\documentclass[fleqn,usenatbib]{mnras}

\usepackage{newtxtext,newtxmath}

\usepackage[T1]{fontenc}
\usepackage{ae,aecompl}

\usepackage{graphicx}	
\usepackage{amsmath}	
\usepackage{amssymb}	
\usepackage{subfig}
\usepackage{xcolor}

\title[HI vs H$\alpha$]
  {HI vs. H$\alpha$ - Comparing the Kinematic Tracers in Modeling the Initial Conditions of the Mice}
\author[Mortazavi et al.]
  {S. Alireza Mortazavi \thanks{mortazavi@jhu.edu}, $^1$
  Jennifer M. Lotz, $^2$
  Joshua E. Barnes, $^{3,4}$
  George C. Privon, $^5$
  \newauthor
  Gregory F. Snyder $^2$\\
  $^1$Department of Physics and Astronomy, Johns Hopkins University, 3400 N. Charles St.,  Baltimore, MD 21218, USA\\
  $^2$Space Telescope Science Institute, 3700 San Martin Dr., Baltimore, MD 21218, USA\\
  $^3$Institute of Astronomy, University of Hawaii, 2680 Woodlawn Drive, Honolulu, HI 96822, USA\\
  $^4$Yukawa Institute for Theoretical Physics, Kyoto University, Kitashirakawa Oiwakecho, Sakyo-ku, Kyoto 606-8502, Japan \\
  $^5$Instituto de Astrof\'isica, Pontificia Universidad Cat\'olica de Chile, Vicu\~na Mackenna 4860, 7820436 Macul, Santiago, Chile}

\date{Accepted XXX. Received YYY; in original form ZZZ}

\pubyear{2016}

\begin{document}
\label{firstpage}
\pagerange{\pageref{firstpage}--\pageref{lastpage}}
\maketitle

\newcommand{\nii}{[N II]}
\newcommand{\ha}{H$\alpha$}
\newcommand{\logniiha}{log$_{10}$(\nii/\ha)}

\begin{abstract}
We explore the effect of using different kinematic tracers (HI and \ha) on reconstructing 
the encounter parameters of the Mice major galaxy merger (NGC 4676A/B). We observed the Mice 
using the SparsePak Integral Field Unit (IFU) on the WIYN telescope, and compared the \ha\ 
velocity map with VLA HI observations. The relatively high spectral resolution of our 
data (R $\approx$ 5000) allows us to resolve more than one kinematic component in the emission lines of some fibers. We separate the \ha-\nii\ emission of 
the star-forming regions from shocks using their \nii/\ha\ 
line ratio and velocity dispersion. 
We show that the velocity of star-forming regions agree with that of the cold gas (HI), 
particularly, in the tidal tails of the system. We reconstruct the morphology and kinematics of these tidal tails utilizing an automated modeling method based on the Identikit software package. We quantify the goodness of fit and the uncertainties 
of the derived encounter parameters. 
Most of the initial conditions reconstructed using \ha\ and HI are consistent with each other, and qualitatively agree with the results of previous works. For example, we find 210$\pm^{50}_{40}$ Myrs, and 
180$\pm^{50}_{40}$ Myrs for the time since pericenter, when modeling 
\ha\ and HI kinematics, respectively. This confirms that in some cases, \ha\ kinematics 
can be used instead of HI kinematics for reconstructing the initial conditions of galaxy mergers, 
and our automated modeling method is applicable to some merging systems.
\end{abstract}




\section{Introduction}

Mergers are key processes in galaxy formation and evolution.
They are one of the major contributors to the mass assembly of galaxies,
they induce starbursts in galaxies, and they are likely to be responsible for the transformation 
of disc-dominated, rotation-supported 
galaxies to bulge-dominated, dispersion supported ones (\citealt{Toomre:1972jia},
\citealt{Barnes:1996bna}, \citealt{Mihos:1996boa}). 

Reconstructing the encounter parameters of galaxy mergers (including their initial conditions, 
orbital parameters, and observer-dependent parameters) via dynamical modeling puts 
new constraints on our understanding of galaxy evolution and cosmology. For example, 
isolated hydrodynamical disc-disc galaxy mergers have shown that the initial conditions 
such as pericentric separation and initial orientation of discs affects the timing and amount
of merger induced star formation (\citealt{Cox:2008jj}, \citealt{Snyder:2011fs}). Dynamical 
modeling constrains these parameters independent of the measured star formation history 
in the system,so one can use them as independent tools for testing models of merger-induced 
star-formation. In addition, we have learned that the 
initial orientation of interacting discs correlates with whether the remnant will be a
 fast or a slow rotator. (\citealt{Bois:2011kc}; \citealt{Naab:2014ht}). Besides, we can
put constraints on cosmological dark matter 
simulations by measuring the orbital parameters (e.g. eccentricity) 
for a statistical sample of galaxy mergers. Distribution of orbital parameters of galaxy mergers can be compared 
with that of dark matter halo mergers in cosmological simulations (e.g. see \citealt{Benson:2005hi},
\citealt{Khochfar:2006de}).

Interacting galaxies often experience strong tidal forces when they pass
by each other, producing pronounced tidal features in discs. These features are usually strongest 
after the first pericenter, though their
shape and strength is a complex function of the initial parameters of the 
orbit and the structure of the galaxies (\citealt{1996ApJ...462..576D}, 
\citealt{Springel:1999hv}, \citealt{Barnes:2016gg}). The sensitivity of the 
shape and velocity of these features to the initial conditions
make them strong tools for modeling the dynamics of galaxy mergers and
their initial conditions \citep{Barnes:2009fh}. 

Dynamical modeling of galaxy mergers is possible by finding the most similar simulation
to the morphology and kinematics the data. Here,``most similar" is 
a vague term. Most previous attempts to model galaxy mergers have used qualitative, 
subjective matching criteria obtained by visual inspection of the model and data 
(e.g. \citealt{Toomre:1972jia};  \citealt{Hibbard:1995iz}; \citealt{Barnes:2009fh}). 
In \cite{Mortazavi:2016hv} we developed an automated method based on Identikit 2 
(\citealt{Barnes:2011kb}). In this method we use collisionless massless particles 
to reproduce tidal features. Our method 
is not only less subjective than the visual matching techniques, but also provides well-defined 
error-bars for the initial parameters.

For modeling a galaxy merger, we need to know the line of sight velocity of tidal features.
There have been some attempts to model merger systems without implementing velocity
information, only trying to match the morphology of the model with data 
(\citealt{Shamir:2013by}, \citealt{Holincheck:2016fy}). Velocity measurement requires
resolved spectroscopy which tends to be more expensive and less available than
imaging data. Optical morphology of many galaxy mergers are/will be available through 
all sky imaging surveys such as Sloan Digital Sky Survey (SDSS, \citealt{York:2000gn}) and 
Large Synoptic Survey Telescope (LSST, \citealt{Abell:2009vz}).
However, there is more degeneracy in the matched solutions when one does not utilize 
velocity information (\citealt{Hibbard:1994kh},\citealt{Barnes:2011kb}), and morphology-only
matching may result in a best-fit model that is inconsistent with the observed velocity 
gradient across the tidal tails and bridges (e.g. \citealt{Borne:1991fo},\citealt{Hibbard:1995iz}). 

Different velocity tracers can be used to measure the kinematics of tidal features.
Velocities of stars are usually ideal to match with test-particle and collisionless self-consistent 
simulations. We may assume that stars have had enough time to redistribute as collisionless particles, 
if the stellar population is formed long before the encounter begins.
Nonetheless, measuring the velocity of stars in the faint tidal tails and bridges  
requires a high signal to noise ratio in the continuum of 
spectra which is expensive to obtain. 
Another option is to measure the velocity of cold neutral hydrogen gas (21 cm HI emission).
Neutral hydrogen is usually more extended than stars in disc galaxies, and 
produce stronger tidal features when discs interact in prograde orbits. Strong tidal features are
useful for constraining the dynamical model. However, one should keep in mind that 
cold gas is dissipative. Some dissipative structures are produced through chaotic processes, 
and it is hard to reproduce them not only with collisionless test particles of Identikit 
\citep{Barnes:2009fh,Mortazavi:2016hv}, but also in hydrodynamical simulations including 
gaseous components. It is also expensive to measure the kinematics of cold
gas in tidally interacting galaxies with a spatial resolution that resolves the velocity 
gradient across the tails.
The third option is to measure the velocity of star forming regions using nebular line emission (e.g. \ha). 
Interaction induce star formation in
gas-rich disc galaxies, and one often finds H II regions in the tidal tails and bridges 
\citep{Jog:1992ct,deGrijs:2003jr}. Measuring
\ha\ emission is a lot less expensive than measuring the velocity of stars or the 
cold gas (HI). Nonetheless, gas dissipation also affect these regions. In addition, before using
\ha\ as velocity tracer, we must make sure that 
the ionized gas resides with the bulk of baryonic matter and is not displaced in 
position or velocity by non-gravitational phenomena such as high velocity shocks 
driven by supernovae (SNe) or active galactic nuclei (AGN). 

Here, we explore the effect of using different kinematic tracers on the reconstructed 
encounter parameters. IFU galaxy surveys like SAMI (\citealt{Croom:2012fo}), 
CALIFA (\citealt{Sanchez:2012ku}), and MaNGA (\citealt{Bundy:2015ft}) 
measure the resolved \ha\ kinematics of a relatively large 
number of galaxies, including major mergers. Consistency between modeling \ha\ kinematics
and modeling the more extended HI emission shows that our method can be applied on the data from 
these large surveys. As a result, we will have dynamical models of statistically 
significant samples, and we can  estimate the distribution of orbital parameters of major galaxy
mergers in the nearby Universe.

In this work, we focus on modeling one of the most famous galaxy merger systems, 
NGC 4676 (a.k.a. Arp 242, the Mice). NGC 4676 is an early stage galaxy merger
at redshift z=0.02205. It has a very distinctive morphology consisting of two strong 
tidal tails in the north and south of the system resembling two playing Mice. The straight tail
in the northern galaxy indicates that this galaxy is almost edge-on. The southern disc seems
to be slightly tilted, though still close to an edge-on view. In this work, we model
the Mice using the kinematics of two different components:
cold gas (HI), and star-forming regions (\ha). In \S 2 we describe reduction and analysis
of the SparsePak IFU data for obtaining the \ha\ kinematics of the Mice. In \S 3 we briefly mention
the characteristics of the HI data. In \S 4 we describe the method we use for reconstructing
the initial conditions of the Mice using both \ha\ and HI line-of-sight velocity maps, and we
present the modeling results. In \S 5 we discuss these results, comparing them to some previous 
reconstructions of encounter parameters of the Mice in the literature and demonstrating some of the their implications.  

\section{SparsePak IFU Data}
\label{sec:sparsepakifudata}

\subsection{\ha-\nii Observations}
\label{sec:observations}

\begin{figure}
 \centering
\includegraphics[width=0.45\textwidth]{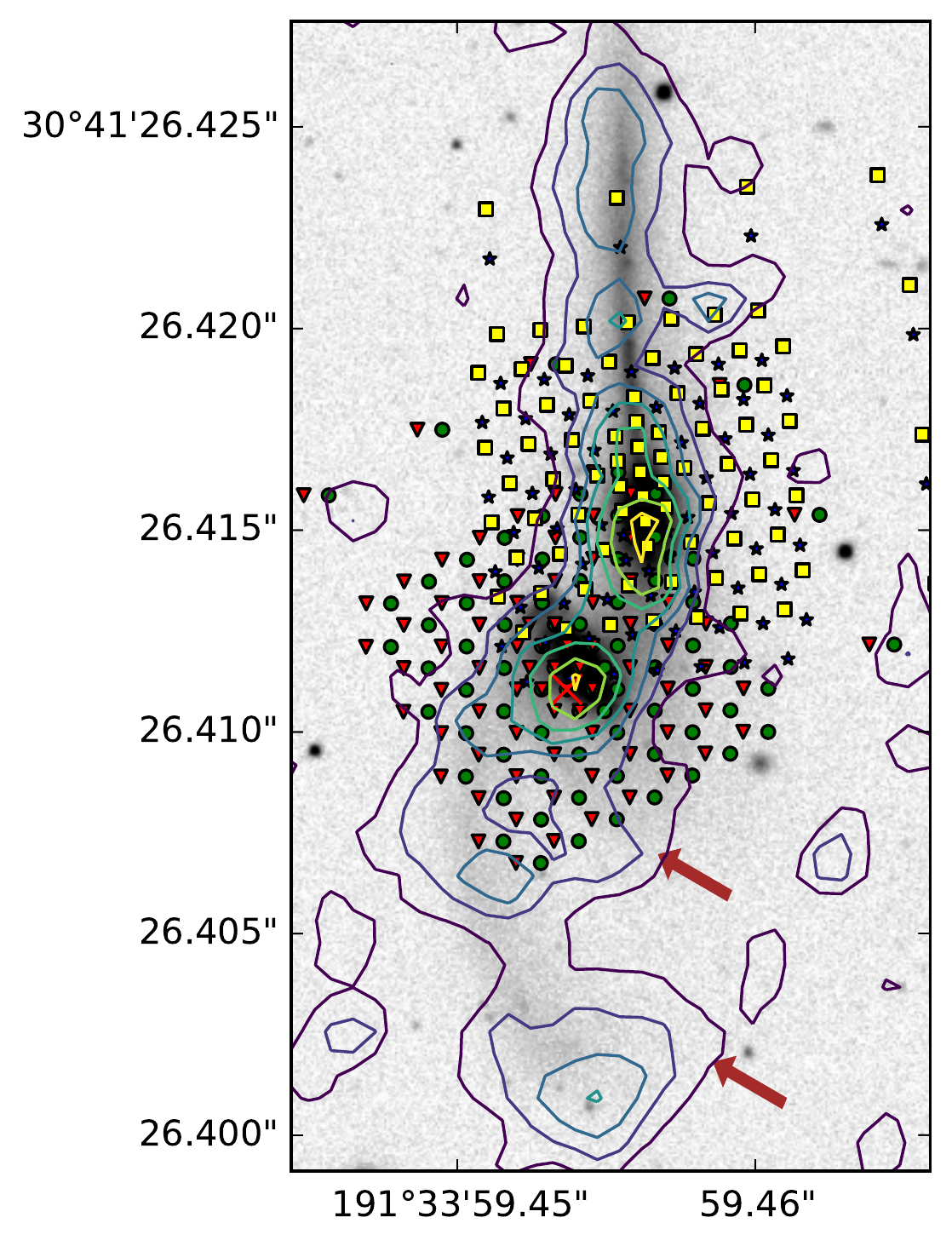}
\caption[The HI surface density contours plotted over SDSS r-band image and SparsePak pointings]
{The HI surface density contours are plotted over the SDSS r-band
image. Sky positions of the fibers in four SparsePak pointings are shown
separately, with green circles, red triangles, blue stars, and yellow squares. 
The red cross indicates the fiber for which the spectrum is shown in Figure \ref{fig:spectrum}.
The centers of the two galaxies and the northern tail match in 
r-band image and HI map. The southern tail, however, appears to have
two self-gravitating clouds of cold gas (indicated by the 
brown arrows). These self-gravitating features can not be
reproduced with test-particles in Identikit. We use the
morphology of the r-band image along with the kinematics of the HI gas.}
\label{fig:matchcoldgas_stars_fibers}
\end{figure}

We observed the Mice using the SparsePak
Integral Field Unit (IFU) on the WIYN telescope at Kitt Peak National Observatory 
(KPNO)(\citealt{Bershady:2004gp}) in 
March 2008. Our goal was to measure the kinematics of \ha\ 
emission line. We do 
not require a uniform coverage of velocity information over the system.
Often, just the velocity of a few fibers in the tail regions is enough to 
break the degeneracy in the merger parameter space. SparsePak is especially suitable for 
this purpose as it has a relatively large field of view ($\sim 1'$), at the expense 
of missing areas between the sparsely placed fibers. We observed the Mice with four SparsePak 
pointings. Figure \ref{fig:matchcoldgas_stars_fibers} shows the layout of the fiber positions on the Mice. 


For SparsePak observations we used  the bench spectrograph and the 860 lines/mm 
grating blazed at 30.9$^{\circ}$ in order 2, obtaining a dispersion of 0.69 \AA/pixel
(FWHM) in the wavelength range of 6050-7000\AA. We obtain a velocity resolution of
$\sim$ 31 km s$^{-1}$. Our spectral coverage is less than 
 recent and ongoing galaxy surveys such as CALIFA, and MaNGA, But our 
 spectral resolution is higher. In the red band, the dispersion of CALIFA, 
 and MaNGA surveys are 2.0 \AA/pixel and 0.83\AA/pixel, respectively.
 Higher spectral resolution enables us to resolve multiple emission line components, 
 usually appearing in the central regions of galaxies where multiple gaseous components 
 overlap. 
 
 \subsection{Data Reduction}
\label{sec:datareduction}
 
\begin{figure}
\centering
\subfloat[]{\label{fig:SparsePakspectrum}\includegraphics[width=0.45\textwidth]{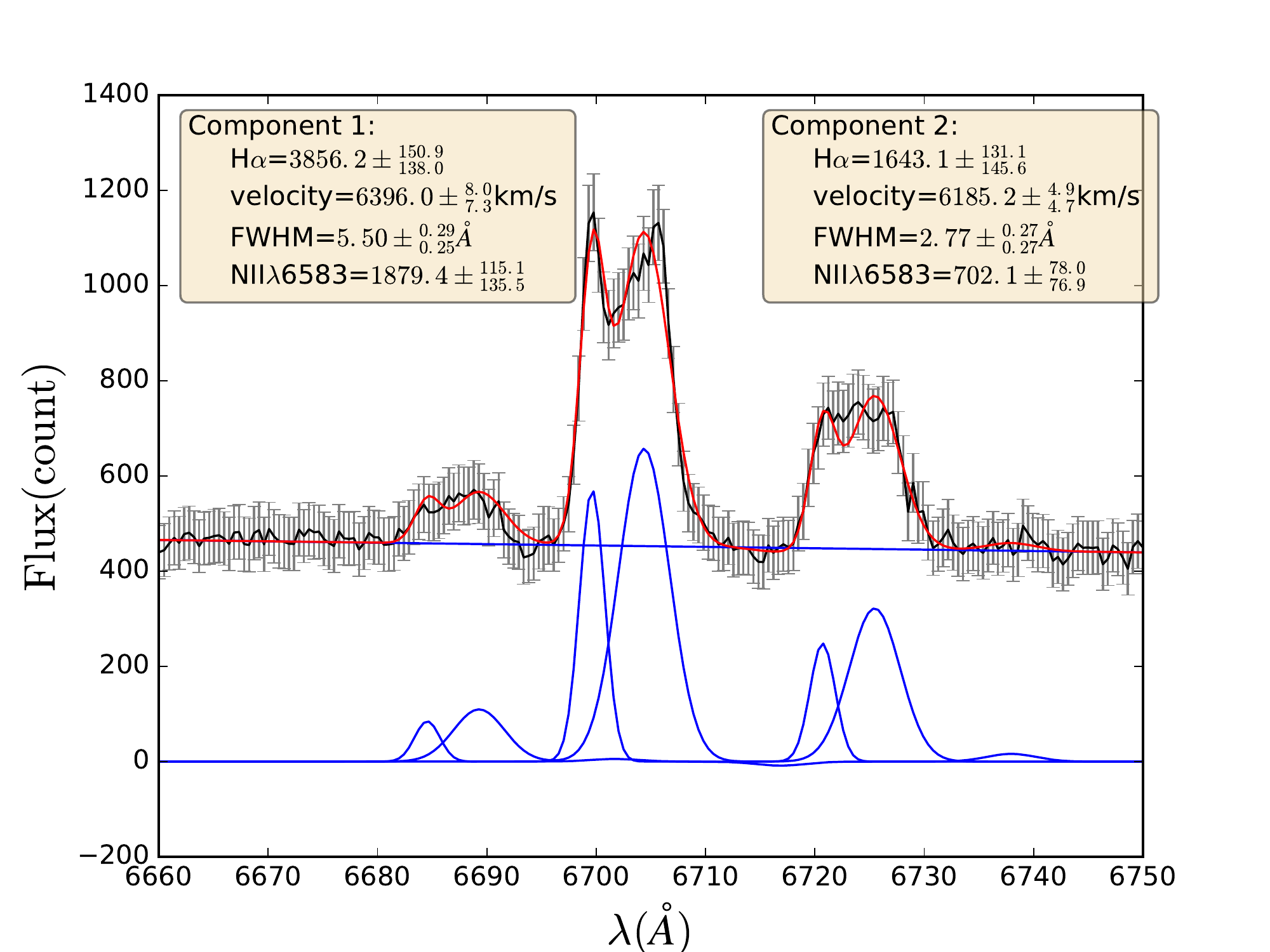}}\\
\subfloat[]{\label{fig:CALIFAspectrum}\includegraphics[width=0.45\textwidth]{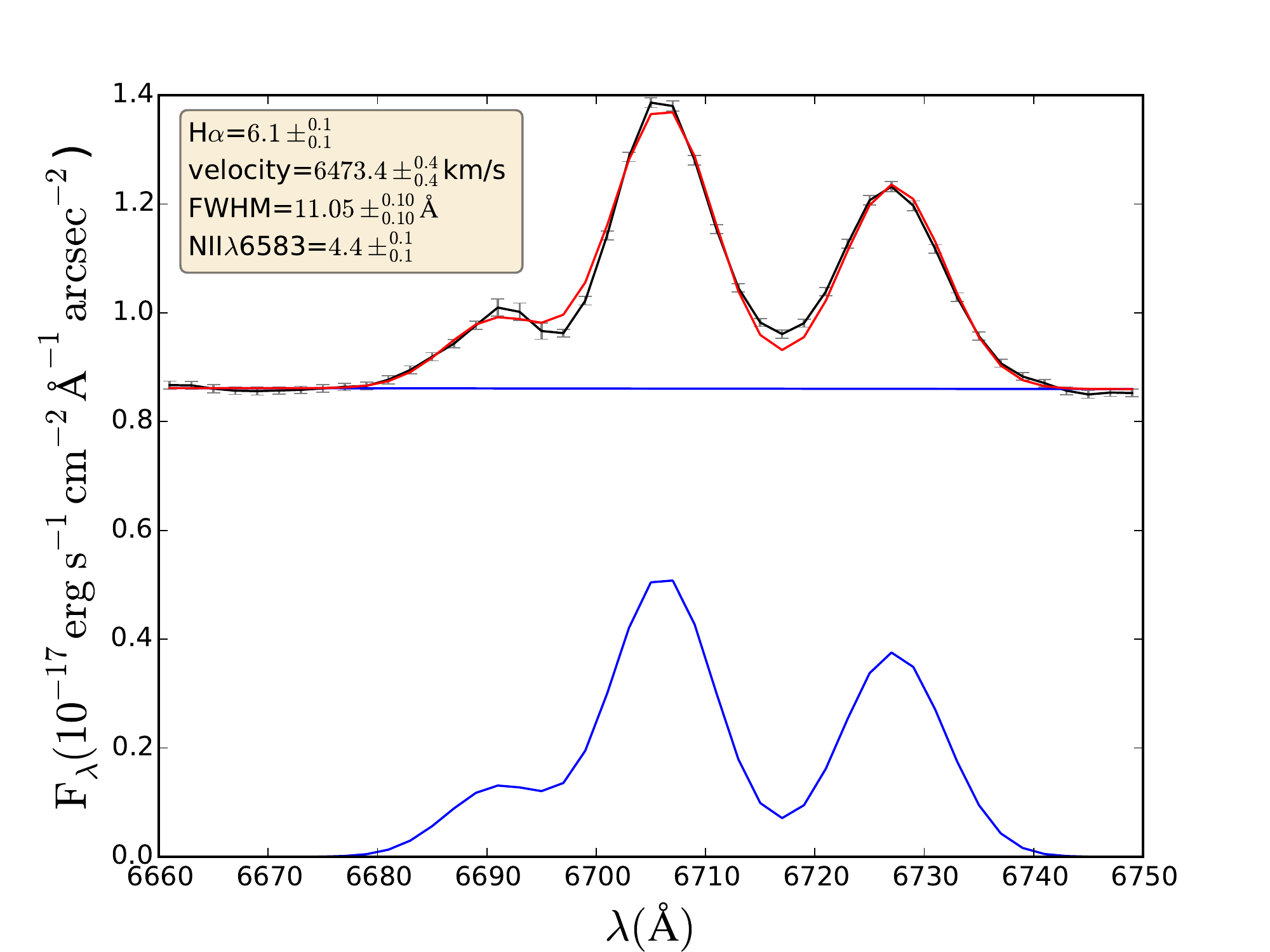}}
\caption[SparsePak spectrum example]
{(a) An example of the SparsePak spectrum in one of the 
central fibers (indicated with the red cross in Figure \protect\ref{fig:matchcoldgas_stars_fibers}). 
The black points with error bars show the spectrum data points. The solid blue curves
show the fitted triple Gaussians (two components) and the the fitted background line.
The red solid line shows the sum of fitted functions. One 
visually confirms that more that one component is 
required to fully describe the emission lines. In this work, we test whether two components are
resolved in each fiber. (b) CALIFA spectrum from the exact same sky position. The points
and solid lines are the same as (a). The lower spectral resolution 
of CALIFA instrument washes out the multiple components.}
\label{fig:spectrum}
\end{figure}

We fit one-component  and two-component triple Gaussian curves on 
top of a line with free slope as the background (continuum) 
over the wavelength range 6665-6755 $\AA$ where \ha\ and \nii\  emission lines 
appear for the Mice system. The triple Gaussian curves have free amplitudes, with centers separated
by the wavelength difference between \nii[$\lambda6548$], \ha[$\lambda6563$], 
and \nii[$\lambda6584$]. The ratio of \nii[$\lambda6584$]/\nii[$\lambda6548$] was fixed
to the theoretical value of 2.95 \citep{Acker:1989vj}. The signal to noise of the continuum is not high
enough to fit the stellar model properly. As a result, in our analysis we did not take the
underlying \ha\ absorption into account. We use an F-test to decide 
which fit to the emission lines is favored by our data. Two-component fits are usually 
preferred in the central regions where the signal to noise is higher 
and parts of the system with different velocities are likely to overlap. Figure \ref{fig:twocomp}
shows fibers in which double component fits were favored.
In each of these fibers, narrow and broad components are determined by their velocity dispersion.
In some of the very central fibers even a two-component fit is not enough to 
properly model the shape of the emission line. Figure \ref{fig:spectrum} shows an example of a fiber where
two-component model is preferred. This fiber is marked with the red cross in Figure \ref{fig:matchcoldgas_stars_fibers}.
In Figure \ref{fig:CALIFAspectrum} we compare the CALIFA spectrum of the exact same location
on the Mice. The multiple components are clearly washed out in CALIFA data due to lower spectral resolution.

\subsection{Shocked vs. Star Forming Regions}
\label{sec:shockedregions}

\ha\ emission originates from ionized gas regardless of how the gas was ionized.
In normal HII regions, atomic hydrogen is photo-ionized by the UV emission from O/B stars.
Photoionization does not significantly affect the overall kinematics of the gas relative to old stars and the 
neutral gas in the vicinity. We expect the velocity obtained from \ha\ emission
of these photo-ionized sources to be relatively similar to the velocity of the bulk of 
the baryons, which is mostly governed by gravitational forces. In dynamical modeling 
of a galaxy merger using collisionless test particles only gravitational effects are to be considered. 
So, we expect \ha\ kinematics of normal HII regions to be ideal for our modeling method.
On the other hand, high speed stellar winds, SN remnants, and feedback from AGN produce shocks and heat up the interstellar gas. Shock-heated gas also
emit \ha, but unlike photo-ionized gas, often its final velocity is
significantly affected by non-gravitational processes.
The kinematics obtained from \ha\ emission of shocked gas
can significantly disagree from that of the bulk of the baryons. Therefore, 
it is important to distinguish emission from the photo-ionized and shocked regions.

One way to separate the star-forming and shocked regions is to utilize the 
Baldwin, Phillips \& Terlevich (BPT) diagnostics diagram which uses [O III]$\lambda$5007/H$\beta$, 
\nii$\lambda$6583/\ha, [S II]$\lambda\lambda$6716,6731/\ha, 
and [O I]$\lambda$6300/\ha\ flux ratios (\citealt{Baldwin:1981ev}, 
\citealt{Kewley:2006gb}). Our SparsePak observations, however, 
were limited to wavelength range of 6050-7000 \AA\ and did not include H$\beta$ 
and [O III] emission lines. We need a different method to distinguish 
shocks from normal star-forming regions.

The shocked 
regions tend to have higher \nii/\ha\ and usually exhibit 
higher velocity dispersion.  \cite{Rich:2011is} showed that the histogram
of velocity dispersion of emission line components in some galaxy mergers reveals a bi-modality. The emission from star forming regions and shock heated gas appears to be responsible for the low and high velocity dispersion modes, respectively. Via further analysis of a larger sample, including interacting systems in various merger stages, \cite{Rich:2014ib} chose a limit of $\sigma<90$ km/s for the velocity dispersion of emission lines from HII and turbulent star forming regions. The components with 
$\sigma>90$ km/s are, hence, considered to be emitted from low velocity shocks. On the other hand, [NII] BPT diagram suggests that when \logniiha<-0.2, the emission is more likely to originate from star forming regions, i.e. below the dashed curve in Figure 4a of \cite{Kewley:2006gb}. Components with larger \logniiha are from either the composite or AGN regions of \nii\ BPT diagram. However, many IFU observations of extended nebular emission in galaxies have suggested that the galaxy-wide shocks are the sources of composite emission, not a mixture of AGN and star formation. \citep{MonrealIbero:2006fi,MonrealIbero:2010ko,Farage2010OPTICALFILAMENTS,Rich:2011is,Rich:2014ib,Rich:2015kf} 
Following the argument that shocked gas has higher velocity dispersion and \nii/\ha, we use a plot of velocity dispersion vs. \logniiha 
 to separate the shocked regions from normal 
star forming ones.

\begin{figure*}
\centering
\subfloat[]{\label{fig:NIIHa_sigma}\includegraphics[width=0.40\textwidth]{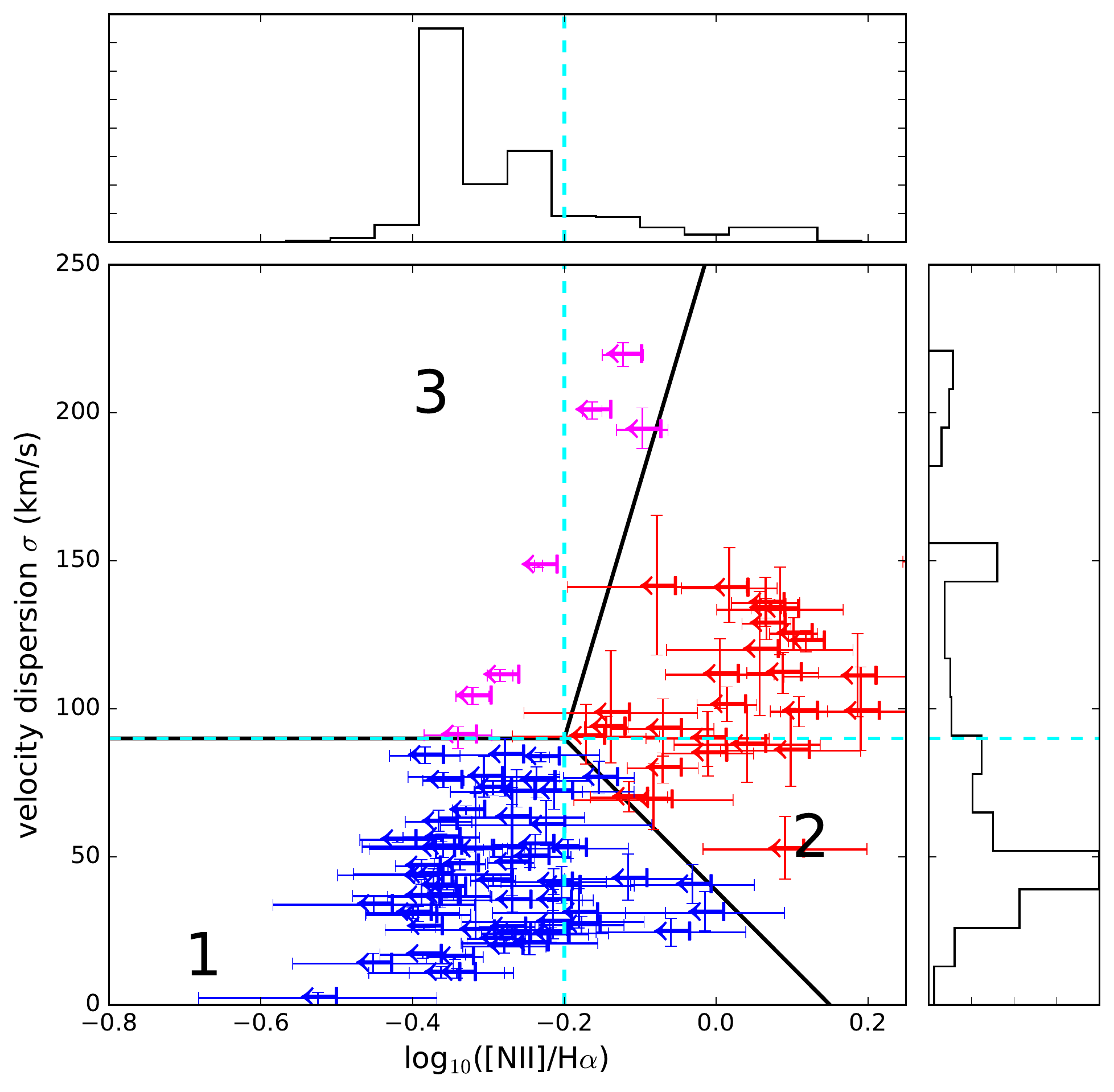}}
\subfloat[]{\label{fig:CALIFABPT}\includegraphics[width=0.40\textwidth]{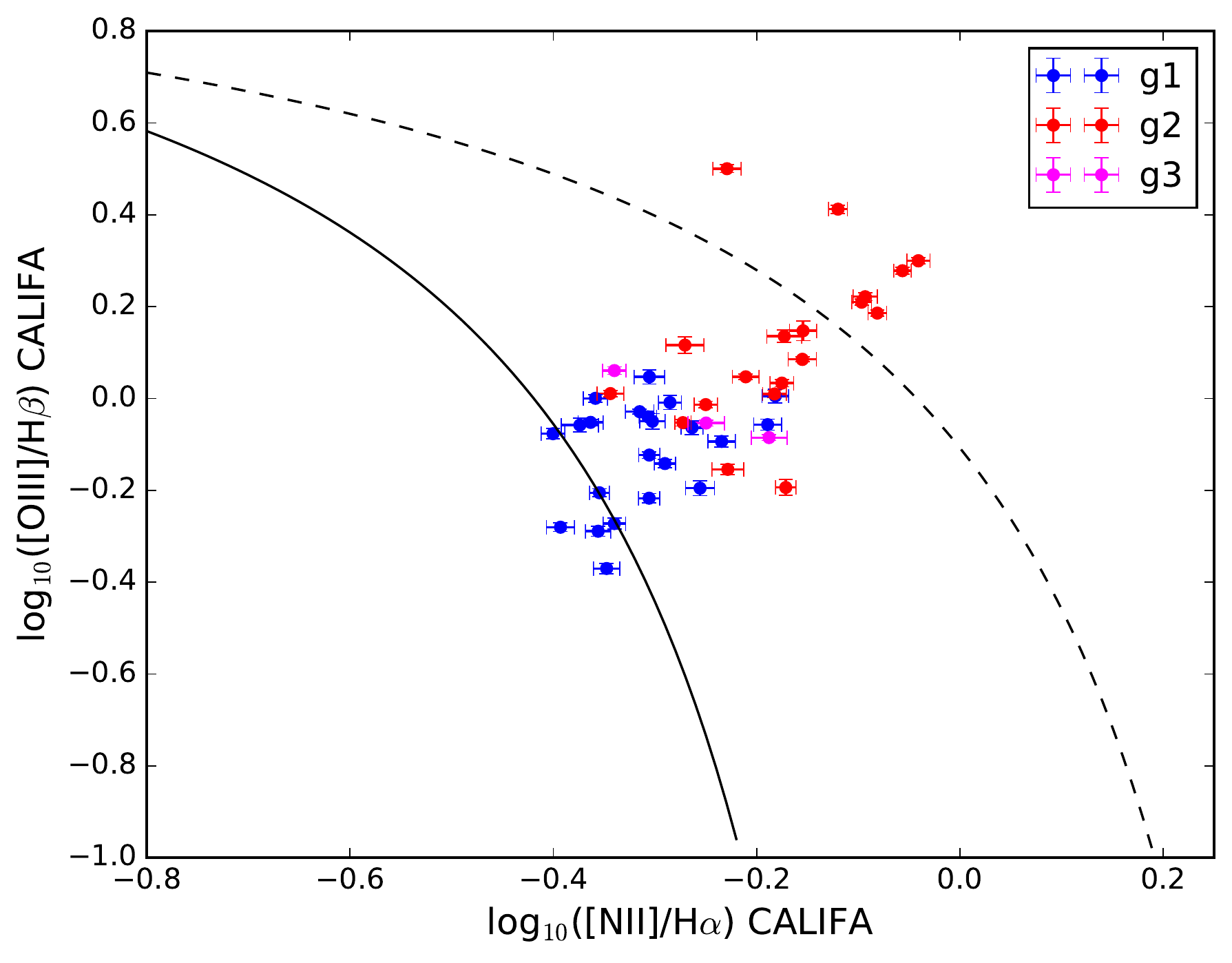}}
\caption[(a) Plot of velocity dispersion vs. \logniiha\ for all components with S/N$>$5. (b) \nii\ BPT diagram for fibers with available CALIFA survey.]
{(a) Plot of velocity dispersion vs. \logniiha\ for all components with S/N$>$5. 
The markers are shown as left-arrows to indicate that the values
of \logniiha\ are upper limit, as we did not take the underlying \ha\ absorption into account in our analysis. Data points are visually classified into 3 groups. 
Group 1 (lower left) are taken as components
emitted from normal star-forming regions. Group 2 components (right) mostly have higher \logniiha\ and
velocity dispersion than group 1. They
are likely to have been emitted from shock-heated gas. Group 3 (upper left) components have large velocity dispersion
and lower \nii/\ha\ compared to group 2, suggesting that they are the projection of multiple unresolved kinematic components.
The horizontal and vertical dashed cyan lines show, respectively, the limits of $\sigma=90$ km/s and
\logniiha$=-0.2$ for separating shocks in the literature. On the top and left panels, the \ha\ flux wighted histogram of these components are shown for \logniiha\ and velocity dispersion, respectively. (b) \nii\ BPT diagnostic diagram for fibers that preferred only one component fit in our data. For this plot CALIFA data has been used. The fibers are color-coded by their group from (a). The solid and dashed lines are from \cite{Kewley:2006gb}, dividing star forming, composite, and AGN regions. Even though most fibers are in the composite region, the location of components in groups 1 and 2 verifies that they are more likely to have been originated at star forming and shocked regions, respectively. Please note that the emission lines are taken from \cite{Sanchez2016CALIFASurvey}, who have modeled the stellar continuum (underlying \ha\ absorption), so they measure a slightly lower value for \logniiha.}
\label{fig:HIIvsshocked}
\end{figure*}

Figure \ref{fig:NIIHa_sigma} shows the plot of velocity dispersion vs. \logniiha\ 
for all kinematic components with S/N$>$5. Through visual inspection, we find that
the data points are clustered into three groups. Group 1 has low 
\nii/\ha\ and low velocity dispersion, 
group 2 has high \nii/\ha\ and high velocity dispersion, and  
group 3 has low \nii/\ha\ and high velocity dispersion.
We chose the black lines shown in 
Figure \ref{fig:NIIHa_sigma} to separate these three groups. Dashed cyan lines show $\sigma = 90$ km/s (limit of shocks in \citealt{Rich:2014ib}) and \logniiha$=-0.2$ (highest value for star forming galaxies in \nii\ BPT diagrams; \citealt{Kewley:2006gb}). On the top and right panels, the \ha\ flux wighted histogram of all components are shown for \logniiha\ and velocity dispersion, respectively. As our data was not flux calibrated we estimated \ha\ flux by the number of electron counts per unit exposure time. A weak bi-modality is visible in the velocity dispersion histogram, similar to systems in \cite{Rich:2011is,Rich:2015kf}; though, the peak of histogram at high velocity dispersion is weaker in our data.

In our spectral 
analysis we did not fit the stellar model to the spectra; instead, 
we used a line with free slope to model
the continuum around \ha-\nii\ triplet. Resultingly, we did not take the 
underlying \ha\ absorption feature of the stellar continuum into account. This means that
The \ha\ equivalent width (EW) is underestimated in our analysis, and the measured \nii/\ha\ are upper limits. 
So, the data points in Figure \ref{fig:NIIHa_sigma} are shown with left arrows.

Group 1 components are taken to be
the components originating from normal star forming regions. We use 
these components to make the \ha\ velocity map for dynamical modeling.
(See Figure \ref{fig:group1}). 
Most of the components in this group are both below and to the left of the 
horizontal and vertical dashed blue lines, respectively. The smooth rotation
seen across circles in Figure \ref{fig:group1}, and their agreement
with the velocity of cold gas are additional evidences that these
components reveal the velocity of star forming regions.

\begin{figure*}
\centering
\subfloat[]{\label{fig:twocomp}\includegraphics[width=0.3\textwidth]{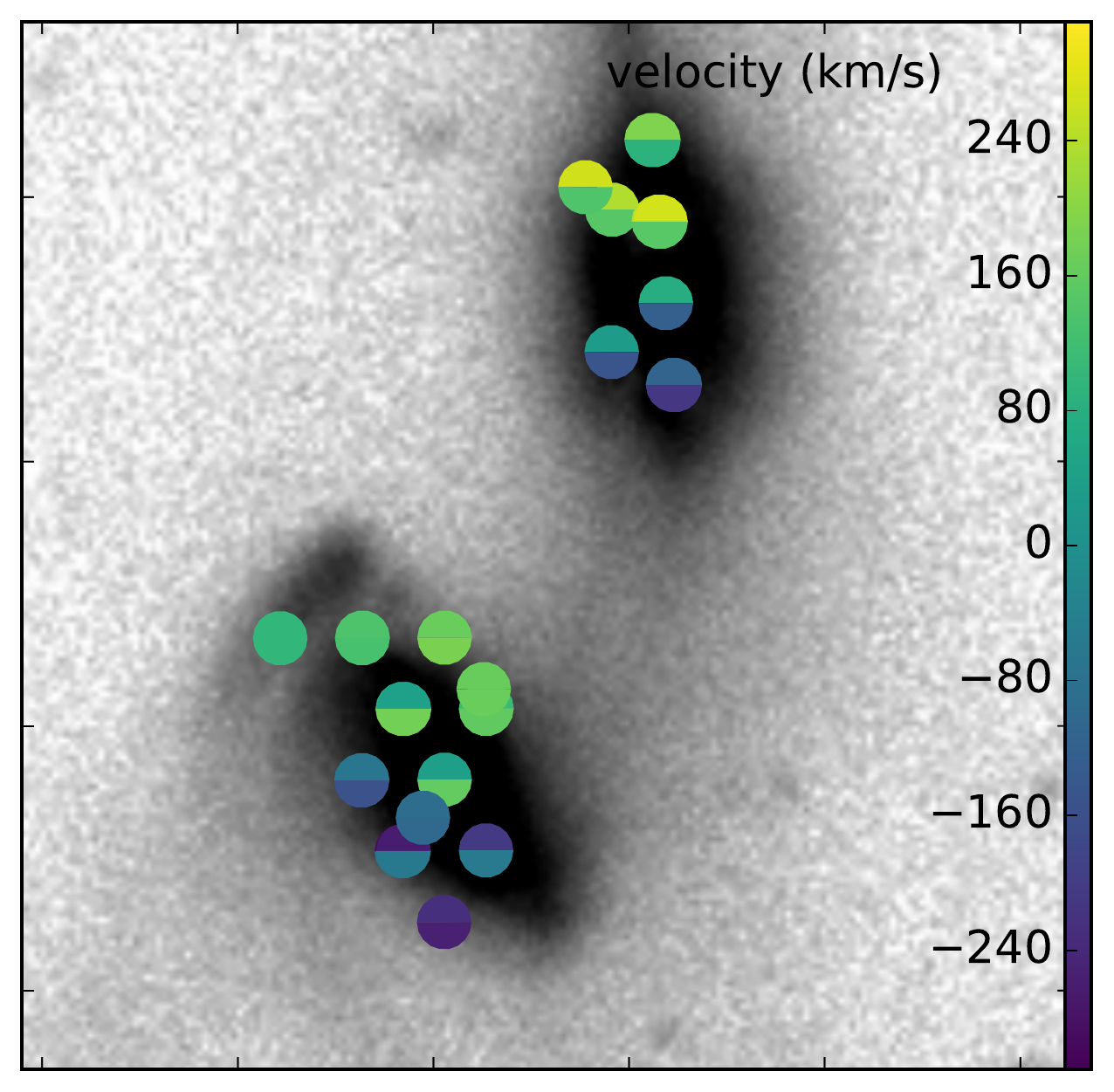}}
\subfloat[]{\label{fig:group2}\includegraphics[width=0.3\textwidth]{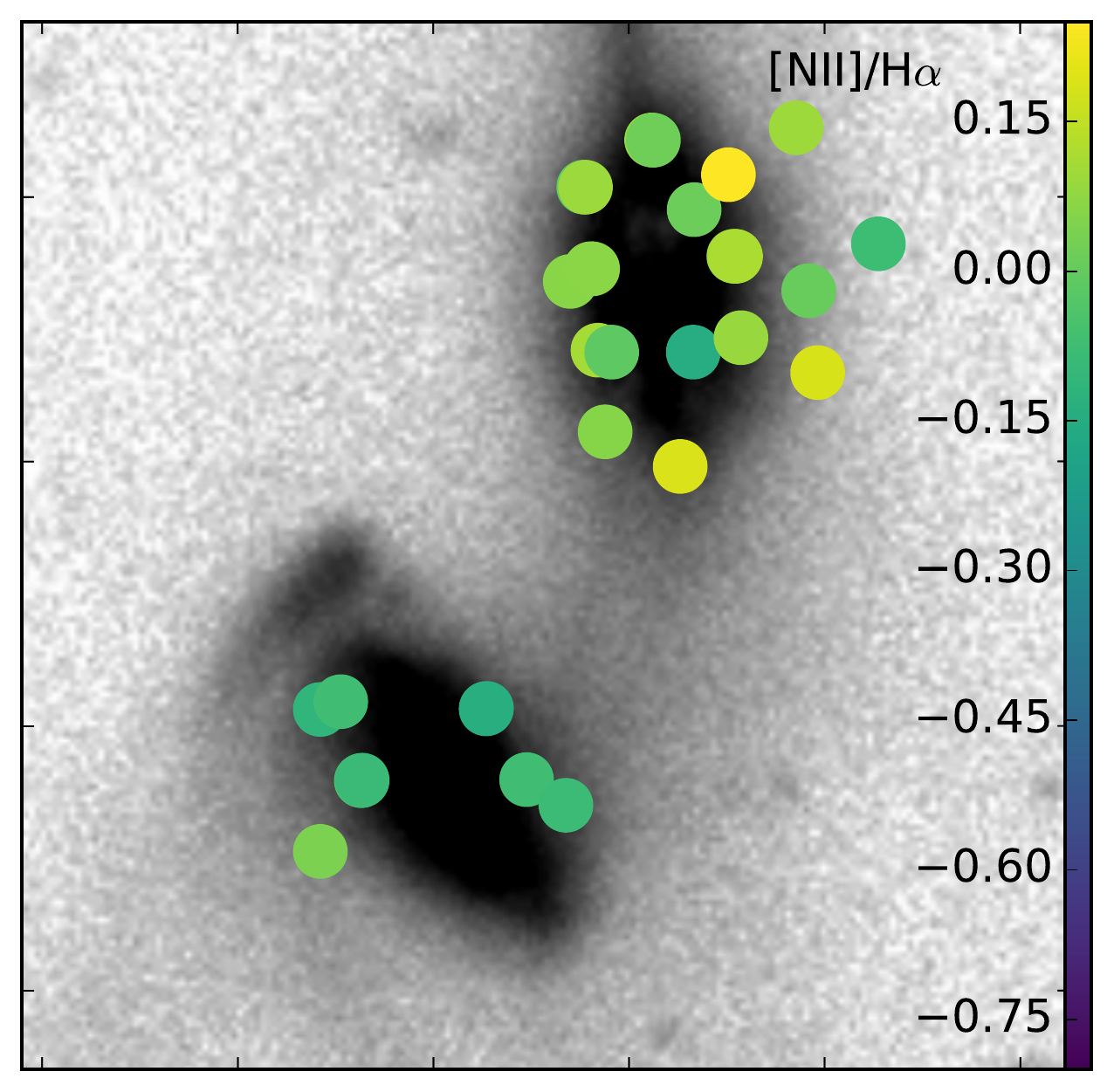}}
\subfloat[]{\label{fig:group3}\includegraphics[width=0.3\textwidth]{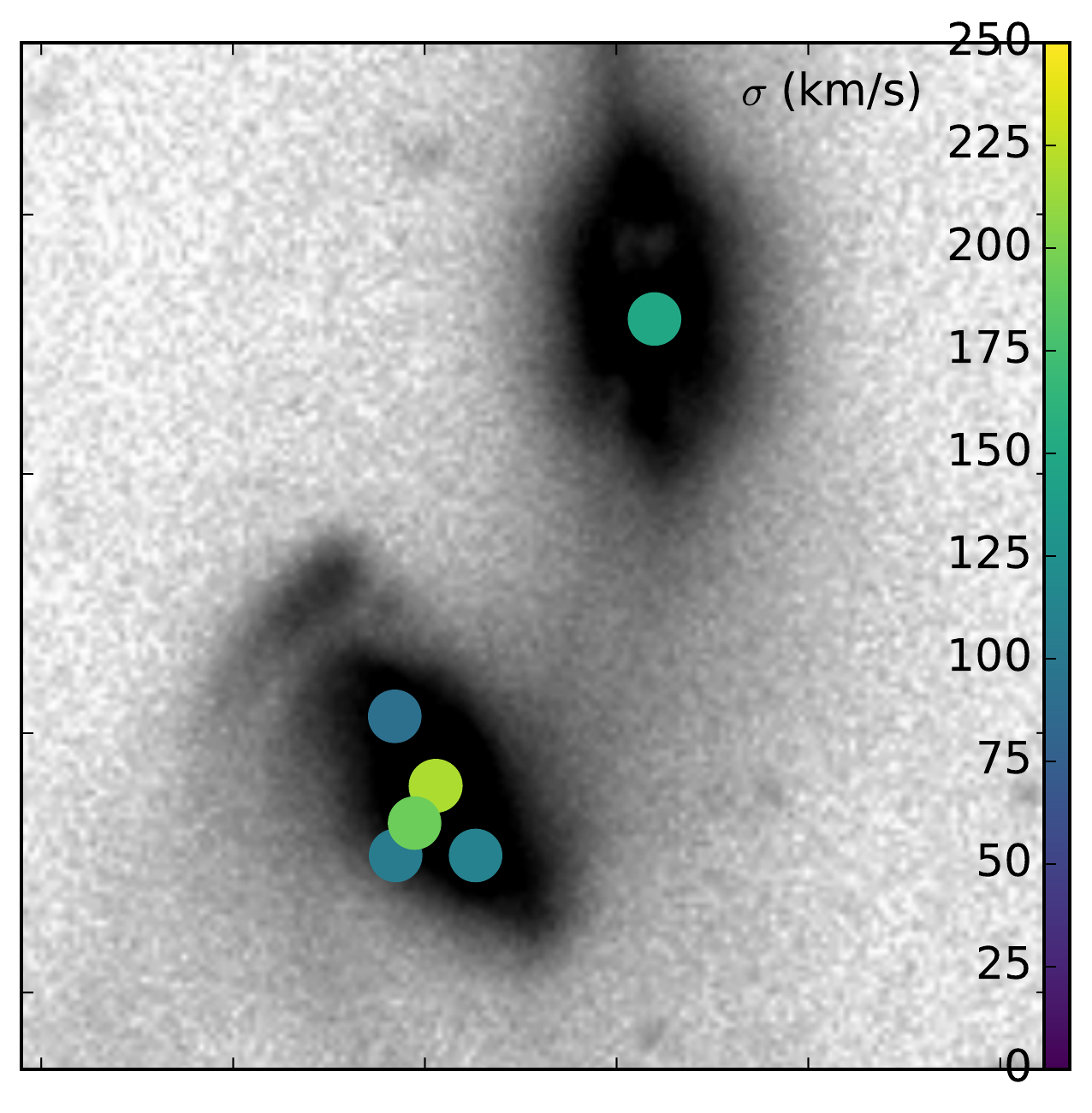}}
\caption[Map of fibers with double components fits, high velocity dispersion and high \nii/\ha]
{
(a) The fibers in which two component fit is preferred over one 
component fit, using the F-test. The velocity of the two components are shown by the color map. 
The upper and lower half of each circle show the velocity of the narrow and broad components, respectively.
(b) The fibers with components in 
group 2 of Figure \ref{fig:NIIHa_sigma}. The spatial distribution of these fibers is 
consistent with the bi-cone shocked structure seen in \cite{Wild:2014do}, and 
suggests that a central process is responsible for shocks. 
(c) The fibers with components in group 3 of Figure \ref{fig:NIIHa_sigma}. These
components have large velocity dispersion and low \nii/\ha. The color of points
show their velocity dispersion. They are close to the central regions where multiple gaseous component are more likely to be projected in 
the same line of sight. Note that the broad component in the spectrum of 
Figure \ref{fig:spectrum} belongs to this group.  }
\label{fig:spatial_distirbutions}
\end{figure*}

Group 2 components are likely to be emitted from the shocked regions ,because they have both higher \nii/\ha\ and higher velocity dispersion. The spatial position
of these components are shown in Figure \ref{fig:group2}. Most of them lie on the two sides, above and below
the discs near the cores of both galaxies. This is consistent with the bi-cones of shocked
material seen in Figure 7 of \cite{Wild:2014do}. \cite{Wild:2014do} used CALIFA data with better spectral coverage, but lower resolution compared to our data. Based on the spatial distribution of [O III]/H$\beta$ ratio and X-ray data, they argue that NGC 4676A do not host an AGN, and the source of ionization is fast outflowing
shocks, probably originating from a starburst in the core. In NGC 4676B, however, they do not rule out the possibility of an 
AGN being the source of bi-conal structure. 
The overall agreement of our data with the spatial 
distribution of regions of high \nii/\ha\ in CALIFA, for which 
stellar model is taken into account, suggests that the underlying \ha\ absorption
does not significantly change our measurements of \logniiha.

The spatial distribution of group 3 components are shown in Figure \ref{fig:group3}. Their high
velocity dispersion and relatively low \logniiha\ can be explained by projection of more than 2 components that are not 
resolved. Figure \ref{fig:group3} provides additional evidence for this, showing that these components are situated in regions close to the center of the two galaxies, where it is more likely to have multiple overlapping kinematic components.

In Figure \ref{fig:CALIFABPT}, we show the distribution of components from these 3 groups on the traditional BPT diagram. Our data does not cover [O III] and H$\beta$ lines, but we can find them in CALIFA survey data. CALIFA survey has a lower spectral resolution than our data, as shown in Figure \ref{fig:spectrum}. We do not see multiple kinematic components in CALIFA emission lines. Also, CALIFA only covered the central regions of the two galaxies, missing the tidal tails that we covered partly. In order to inspect the components of Figure \ref{fig:NIIHa_sigma} in BPT diagram using CALIFA data, we select SparsePak fibers in CALIFA footprint in which one-component fit was preferred. We use the CALIFA emission line maps by \cite{SanchezMenguiano:2016uw}, calculate the sum of emission lines of spexels within each fiber, and measure a CALIFA line ratio for each SparsePak fiber. Figure \ref{fig:CALIFABPT} shows that most fibers are in the composite region, shown by the solid and dashed lines from \cite{Kewley:2006gb}. However, fibers in groups 1 and 2 appear to be spread along a mixing sequence from HI region to AGN. Fibers in group 1 tend to have a more HII region-like emission, while group 2 clearly indicates a harder source of ionization. This provides additional support for our method of separating shocks from star forming regions, only using velocity dispersion and \logniiha. Group 3 components are situated in the middle of the other two groups, in line with our suggestion that they could be unresolved combination of group 1 or 2 components.

\begin{figure*}
\centering
\includegraphics[width=0.95\textwidth]{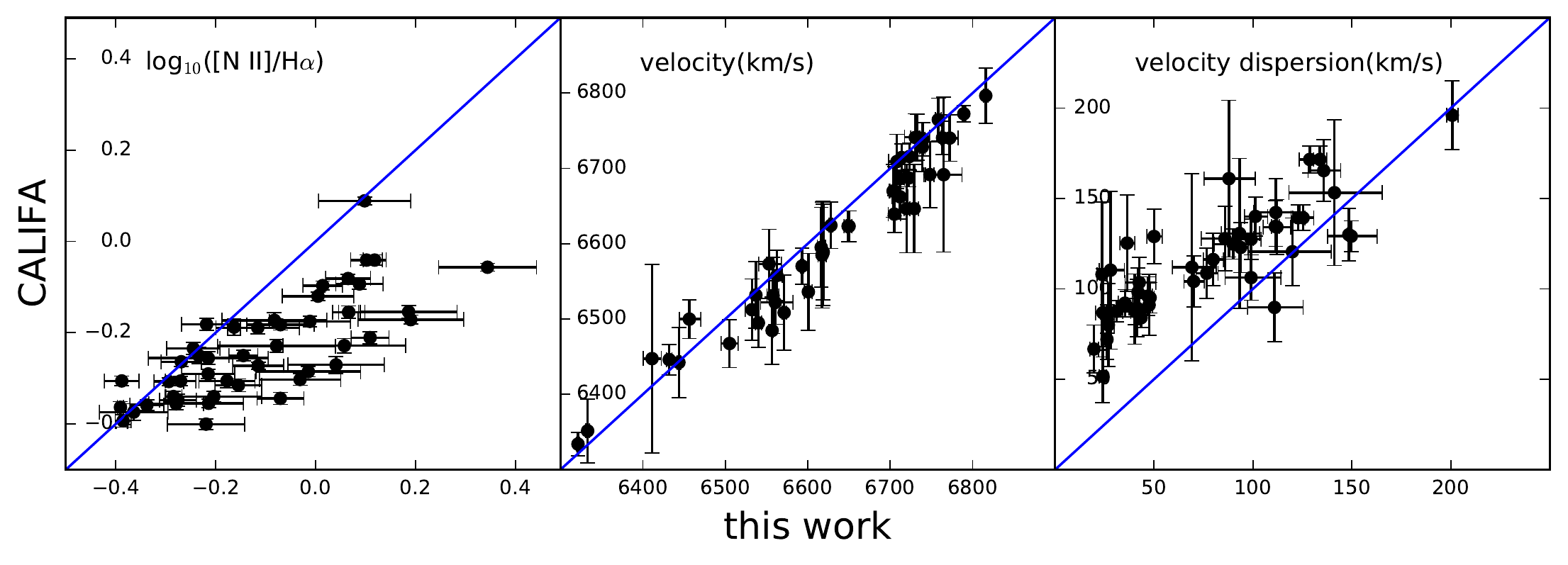}
\caption[Plot of values of \logniiha, velocity, and velocity dispersion derived from our data vs. CALIFA data]{Plot of values of \logniiha, velocity, and velocity dispersion derived from our data vs. CALIFA data \citep{Sanchez2016CALIFASurvey}. Here we only show fibers in which one-component fit is preferred in our high resolution data. In the left panel, \logniiha\ from our data is slightly higher that that of CALIFA, which is the result of the lack of stellar model in our emission line models. The velocities shown in the middle panel agree within one $\sigma$. Velocity dispersion measured in this work is generally smaller, specially for fibers with low velocity dispersion. This is due to the better spectral resolution of our data.}
\label{fig:CALIFAvsSparsePak}
\end{figure*}

\begin{figure*}
\centering
\subfloat[]{\label{fig:group1}\includegraphics[height=0.5\textwidth]{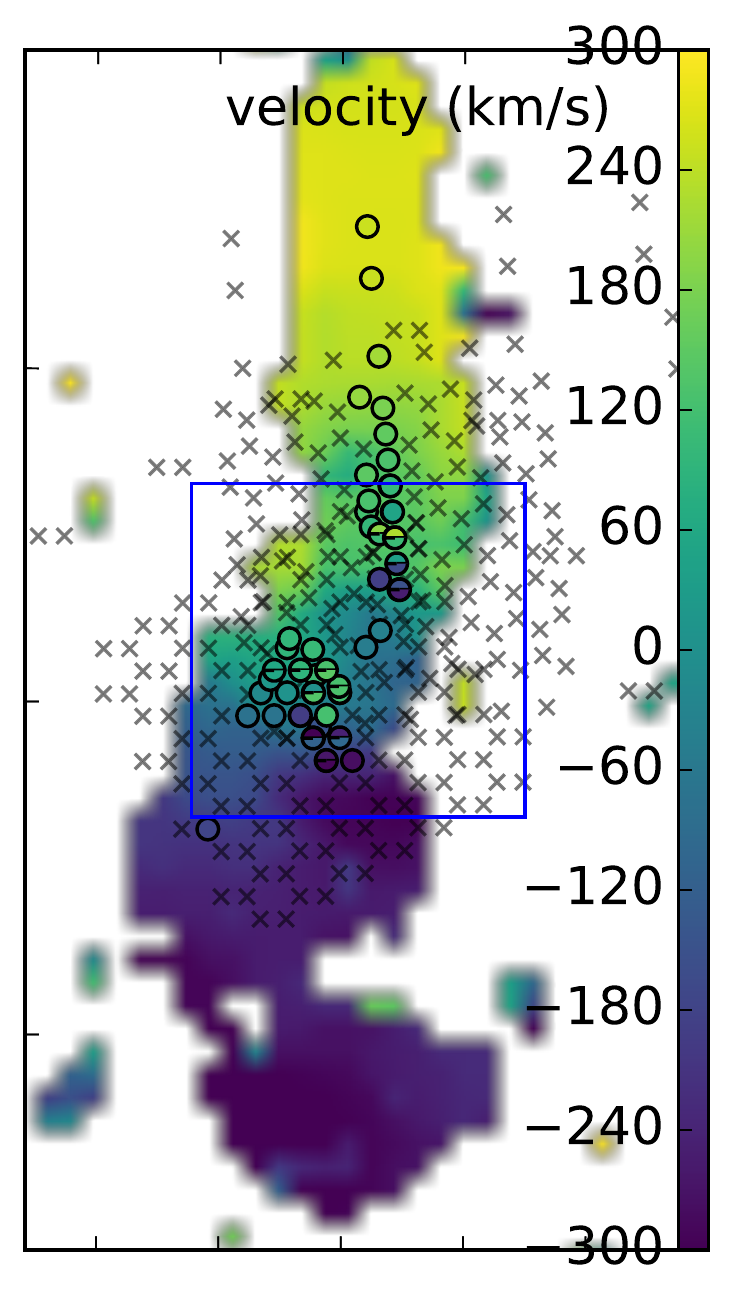}}
\subfloat[]{\label{fig:HIvsHavel}\includegraphics[height=0.5\textwidth]{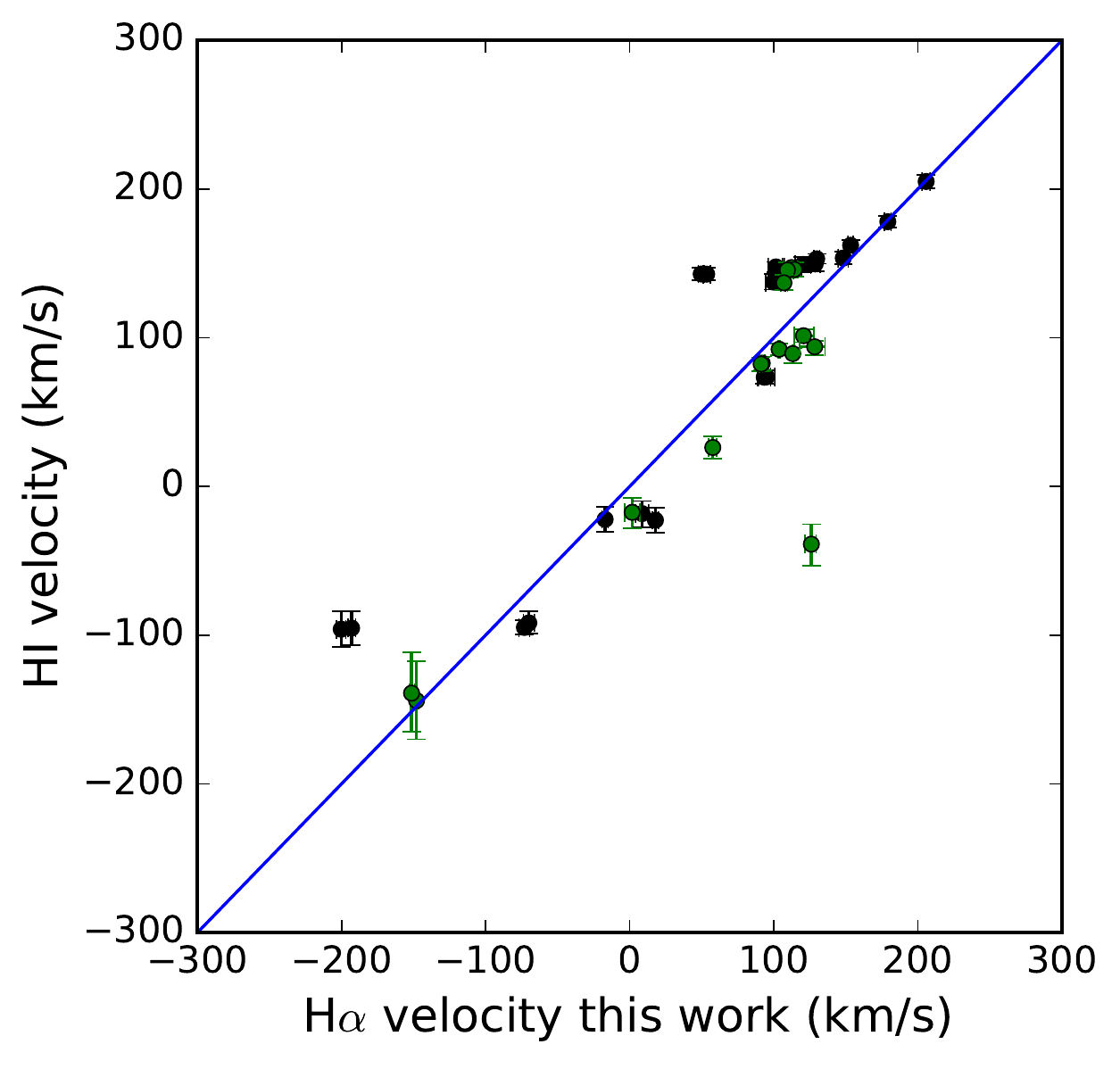}}
\caption[HI vs. group 1 \ha\ velocity]{(a) Velocity of components in group 1 of Figure \ref{fig:NIIHa_sigma}, which have low velocity
dispersion and low \nii/\ha, over-plotted on the
HI Velocity map. There is an agreement in the overall velocity gradient and the 
velocity of tidal tails. The blue box shows the zoom-in area of panels in Figure \ref{fig:spatial_distirbutions}. (b) The velocity obtained from HI vs. group 1 \ha\ emission lines at the location of corresponding fibers. Some HI and \ha\ emission lines reveal two kinematic components. The black data points are for the fibers where a single component fit is preferred for both HI and \ha. The green points are for fibers in which either HI or \ha\ preferred a double component fit, and we only plot the velocities of the components that better match to the other velocity tracer.}
\end{figure*}

Careful reader may notice that in Figure \ref{fig:CALIFABPT} some fibers in group 2 have \logniiha<-0.2, while in Figure \ref{fig:NIIHa_sigma} group 2 components are defined to have a minimum value of -0.2 for \logniiha. This inconsistency is due to the fact that the model of emission line ratios of Figure \ref{fig:CALIFABPT} by \cite{SanchezMenguiano:2016uw}, includes stellar continuum model and takes the underlying \ha\ absorption line into account. As a result, the \logniiha\ at a given fiber of Figure \ref{fig:NIIHa_sigma} tend to be slightly smaller in Figure \ref{fig:CALIFABPT}. This is shown, more clearly, in the left panel of Figure \ref{fig:CALIFAvsSparsePak}. Figure \ref{fig:CALIFAvsSparsePak} shows \logniiha, velocity, and velocity dispersion derived from our data vs. CALIFA data \citep{Sanchez2016CALIFASurvey} for fibers in Figure \ref{fig:CALIFABPT}. The velocity of the two data sets agree, and the velocity dispersion measured from our data is smaller than that of CALIFA due to the better spectral resolution of our data.

Using these groups, we can estimate the fraction of total \ha\ flux 
emitted from shocked gas. To do so, we divide the sum of the \ha\ flux
from group 2 components to the total \ha\ flux. Group 3 components can be 
unresolved components of both groups 1 and 2. We use \nii/\ha$=-0.2$ limit to separate
them for the purpose of estimating shocked gas fraction. We obtain a fraction of $23\pm1\%$ of \ha\ flux 
being emitted from shock-heated gas. The error of this fraction is obtained from
the variance of shock fraction in 100 bootstrapped samples of data points in
velocity dispersion vs. \logniiha\ space (i.e. Figure \ref{fig:NIIHa_sigma}).  

\section{JVLA HI data}
\label{sec:kinematicsofcoldgas}

The kinematics of cold gas are available from radio interferometric
observations of HI 21 cm emission line performed by \cite{Hibbard:1996do}. The observations took place 
in May of 1991 and May of 1992 using the D and C configurations of then Very Large Array (VLA), respectively. 
The HI data has a spatial resolution of
20'' ($\approx$ 9 kpc) and velocity resolution of 43.1 km/s. The total HI mass in the system is $7.5 \times
10^9 M_{\odot}$ (\citealt{Hibbard:1996do}; \citealt{Wild:2014do}). 

In the Mice, there are a couple of self-gravitating/dissipative features in the southern tail with no
counterparts in the optical r-band image. They are indicated by the brown arrows in Figure 
\ref{fig:matchcoldgas_stars_fibers}. After a few unsuccessful trials to reproduce these features
with Identikit, we decided to use the 
morphology of the optical image (SDSS, r-band) along with the velocity of the HI gas, same as \cite{Privon:2013fs}. 

\subsection{Kinematics of \ha\ vs. H I}
\label{sec:kinematicsofhalphavshi}

Figure \ref{fig:group1} shows the \ha\ velocity map 
the mice galaxies, obtained from group 1 star forming components of Figure \ref{fig:NIIHa_sigma}, over-plotted on the H I velocity map. HI velocities are measured as the average of velocities of channels in the HI data cubes, weighted by flux density at each channel. In this color plots the striking agreement between the 
two velocity maps is evident. This agreement is especially important in the tidal tails which 
play the most important role in dynamical modeling, because we try to find the model that best
matches the morphology and kinematics of these features.  

In order to better show the agreement between velocities of star forming \ha\ emission and the cold gas, in Figure \ref{fig:HIvsHavel}, we show a plot of group 1 \ha\ velocities vs. HI velocities at the location of the corresponding fibers. We noticed that some HI emission lines also reveal more than one kinematic components, so similar to our procedure for \ha-\nii\ triplet, we performed one- and two-component Gaussian fitting on HI 21cm emission line, and used the F-test to determine which fit is preferred. In Figure \ref{fig:HIvsHavel}, black data points show fiber in which both HI and \ha\ data prefer a single component fit. The green points show places that either HI or \ha\ data prefer two components, and we only show the components for which the velocities match. As can be seen in Figures \ref{fig:group1} and \ref{fig:HIvsHavel}, in most fibers, HI and group 1 \ha\ velocities match to within the error bars. The disagreement in some fibers near the center can be attributed to the low spacial resolution of HI data and the beam smearing effect.
\section{Dynamical Modeling}
\label{sec:model}

For dynamical modeling of the Mice, we use the automated pipeline that was developed
based on Identikit (\citealt{Barnes:2009fh}, \citealt{Barnes:2011kb}; \citealt{Mortazavi:2016hv}).
Identikit is a software package for modeling the initial condition of major galaxy mergers. It uses 
a precompiled library of N body simulations of galaxy mergers with different initial conditions.
The discs are modeled by test particles. This facilitates the run of multiple discs at the same time, and 
improves the pace of search in the parameter space.
Interactive visual matching with Identikit was used by \cite{Privon:2013fs} to model the Mice
along with 3 other systems. 

The isolated galaxies consist of collisionless massive and massless test particles. massive 
particles are distributed in spherically symmetric fashion representing the mass of the dark
matter halo, the disc, and the bulge. In these models, the luminous mass fraction, 
$f_L = (M_d + M_b) / (M_h+M_d+M_b)$ is 0.2 and the concentration parameter of the halo mass profile, $c_h$ 
is 4.  The discs are represented by test particles that are initially in circular orbits.
The scale length of the disc , $a_d$, is 1/3 of the scale radius of the dark matter halo, $a_h$.
 \cite{Barnes:2016gg} explored the effect of different structure parameters of the isolated
 galaxies on the morphology of tidal features. In this work we keep working on a single mass model 
 for the sake of simplicity.
 
We use an automated matching 
feature that was introduced in Identikit II \citep{Barnes:2011kb}. In Identikit II, the similarity
between model and data is quantified with an informal measure called score. The user 
places small boxes in the same position as the tidal tails and bridges of the system,
extended in two spatial directions and the LOS velocity direction. Identikit
calculates the scores based on the number of disc test particles residing in these phase
space boxes. The models that better reproduce the tidal
features would place more particles in these boxes, and they would gain a better
score. Identikit II has been applied with a limited range of parameters on
the Mice galaxies by \cite{Barnes:2013tn}. They inspected the models with high score visually
and confirmed that they show good, or fair visual match with the HI data. 

In \cite{Mortazavi:2016hv} we described the method that we have developed for 
automated searching of parameter space and reconstructing the initial parameters with robust 
uncertainties. In this method we first put the boxes on tidal features of the
galaxy merger using a semi-automated technique.
The boxes are placed randomly in regions that are far enough from the center of galaxies. 
Boxes near the center of the galaxies, where test particles have higher number density, are 
always populated, and they do not put strong constraint on the initial conditions of the merger. 
As a result, we put  the boxes over the ends of the tidal tails to enforce the similarity in the overall shape
and the velocity gradient. We ignore self-gravitating features like
stellar clusters or blobs of gas, because test particles in Identikit do not reproduce them.
In cases where no velocity information is available (e.g. the absence of \ha\ in the southern 
tail of the Mice), we put boxes that are only extended in spatial directions, and do not put constraint
on velocity. The size of the boxes are selected to match the spatial
resolution of the observation. In case of the HI data The boxes are 8 kpc
wide, almost matching the spatial resolution of JVLA observation of \cite{Hibbard:1996do}. When modeling
the \ha\ kinematics, the size of the boxes where set to be close to the size of SparsePak fibers
(4 kpc). In places where the error in \ha\ velocity measurement was larger that the extent 
of the box in velocity direction, we increased the size of the box proportionally. Figure \ref{fig:model}
shows a sample of selected boxes in morphology and velocity directions. 

Table \ref{tab:Identikit} shows the range of encounter parameters we explored in this work. 
We produced a library of Identikit models with
a variety of eccentricity, pericentric separation, time since first pass, velocity scaling and 
velocity offset. For future references we call these five parameters ``external parameters". 
Calculating the score for each of these models, Identikit constrains viewing angle, initial 
orientation of discs, length scaling and position offset. We call these parameters ``internal 
parameters". By calculating the score for all library members, we obtain a 5 dimensional score 
map for external parameters. Each point of this score map corresponds to a set of calculated 
internal parameters. The model with the maximum score is the best-fit model and the 
corresponding external and internal parameters are the best-fit encounter parameters.

\begin{table*}
\centering
\begin{tabular}{lll}
	\hline
	Parameter Class		&	Parameter							& Range Tested							\\ \hline
	orbital parameters 		&	eccentricity  						& [0.80-1.10] 								\\
						&	pericentric distance					& [0.03125-1.0000]$\times R_{vir}$ 				\\ 
						&	mass ratio							& 1 										\\ \hline
	observer dependent 		&	time since pericenter					& from first pass to second pass				\\
	parameters			&	viewing angle						& viewed from 320 evenly distributed				\\
						&									& directions on a sphere						\\
						&	position							& set by locking the centers					\\
						&	length scaling $\mathcal{L}$			& set by viewing angle and locking the centers		\\
						&	velocity offset						& [-0.1,0.1]								\\
						&	velocity scaling $\mathcal{V}$			& [-0.500-+0.500]$^*$  						\\ \hline
	initial orientation		&									& 1280 evenly distributed orientations				\\
	of discs				&									&										\\ \hline
\end{tabular}
\caption[Range of encounter parameters explored]
{Range of encounter parameters explored. 
      The total number of parameter space points for which we calculated score
      is $\sim3.2\times10^{10}$.
      $^*$ The velocity (length) scaling, $\mathcal{V}$ ($\mathcal{L}$), relates the dimensionless velocity (length) in Identikit to the 
      physical velocity (length). }
\label{tab:Identikit}
\end{table*}

The uncertainty of the best-fit parameters are calculated using a Bootstrap statistical 
method. Score, by itself, does not provide statistical probability as other measures of goodness of the fit like $\chi^2$.
So, in order to estimate the uncertainty of the scores we re-do the random box positioning procedure several times (10 times in this work), and calculate the score each time.
This process randomly moves the position of boxes on the tidal tails and bridges of the
system. The variation in the scores for each model measures the error of the score, which
eventually, translates into the uncertainty of the best-fit parameters. The models with scores
that are within 1, 2, or 3 standard deviation from the best score, respectively, determine the extent 
of 1$\sigma$, 2$\sigma$, or 3$\sigma$ separation in the parameter space from the best-fit model.

In \cite{Mortazavi:2016hv} we applied our modeling method on
on edge-on views of hydrodynamical simulations of prograde disc-disc interactions (similar to the Mice).
We evaluated the bias in the reconstructed interaction parameters comparing them to the known
input parameters of the hydrodynamical simulations.
We showed that when testing young stars in the simulation (representing the \ha\ data), the average 
bias is not significant in eccentricity, $\mathrm{R_{peri}}$, and merger stage ($<1\sigma$). When testing
the cold gas of the same simulation, the reconstructed physical $\mathrm{R_{peri}}$ was almost 
twice the correct value, but the reconstruction of eccentricity and time was only biased by less that 1$\sigma$.

\subsection{Summary of Results}
\label{sec:results}

 \begin{table*}
\centering
\begin{tabular}{ccccccccc}
\hline
 	&  	fractional 					&	physical 				& 	$e$					& 		merger stage 			&	physical time			&	$(\theta,\phi)$						& 	 NGC 4676A						& 	 NGC 4676B	 					\\
	&	$\mathrm{R_{peri}}$			&	$\mathrm{R_{peri}}$ 		&						&	$\Delta T/T_{\textrm{first to}}$ 	&	 (Myr)				&	degrees$^\circ$					&	$(i_1,\omega_1)$					&	$(i_2,\omega_2)$					\\  
	&	($\mathrm{R_{vir}}$)					&	(kpc)					&						& 	${_{\textrm{second pass}}}^*$	&						&									&	degrees$^\circ$					&	degrees$^\circ$					\\ \hline \hline
\ha	&	$0.38\pm^{0.16}_{0.09}$		&	$17 \pm^{12}_{6}$		&	$\leq0.80\pm^{0.05}_{---}$	&	$0.19\pm^{0.21}_{0.01} $		&	$200\pm^{40}_{30} $		&	$(76\pm^{3}_{9}, 0\pm^{61}_{41})^*$		&	$(21\pm^{11}_{5}, 29\pm^{33}_{38})$	&	$(38\pm^{21}_{10}, 187\pm^{20}_{24})$	\\ \\
HI	&	$0.50\pm^{0.03}_{0.44}$		&	$24 \pm^{8}_{13}$		&	$\leq0.80\pm^{0.15}_{---}$	&	$0.14\pm^{0.70}_{0.01} $		&	$170\pm^{90}_{50} $		&	$(67\pm^{4}_{9}, 52\pm^{9}_{66})$		&	$(56\pm^{31}_{24}, 50\pm^{55}_{96})$	&	$(69\pm^{23}_{27}, 162\pm^{25}_{61})$	\\ 
\end{tabular}
\caption[Reconstructed parameters from modeling \ha\ and HI]
{Reconstructed parameters from modeling \ha\ and HI kinematics. $^*$ The 
resolution of automated search in viewing angle was $\sim 11^\circ$ 
(viewdepth parameter = 3 in Identikit; see \citealt{Barnes:2011kb}). 
The errors reported here are from the width of distribution of viewing 
angles in models with high score. Note that negative azimuthal angles
should be interpreted as a 360$^\circ$ rotation. These parameters match the
definitions of \cite{Privon:2013fs} and can be compared with their values.}
\label{tab:results}
\end{table*}

\begin{figure*}
\subfloat[]{\label{fig:bestfitcuthalpha}\includegraphics[width=0.90\textwidth]{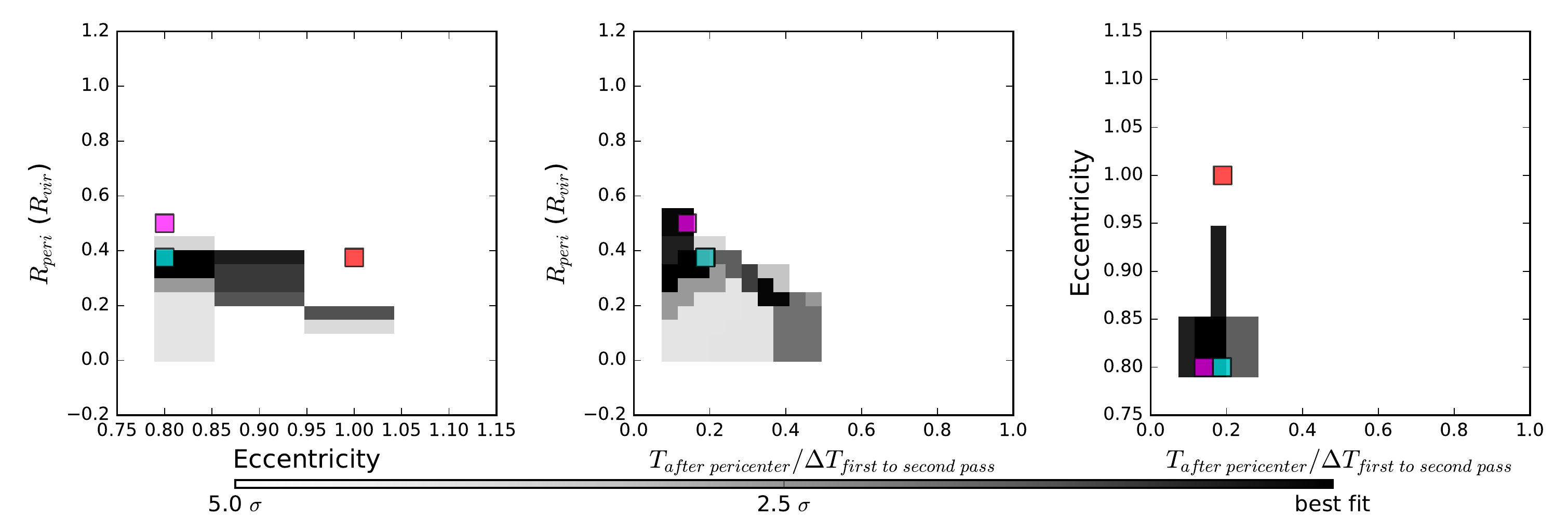}}\\
\subfloat[]{\label{fig:bestfitcuthi}\includegraphics[width=0.90\textwidth]{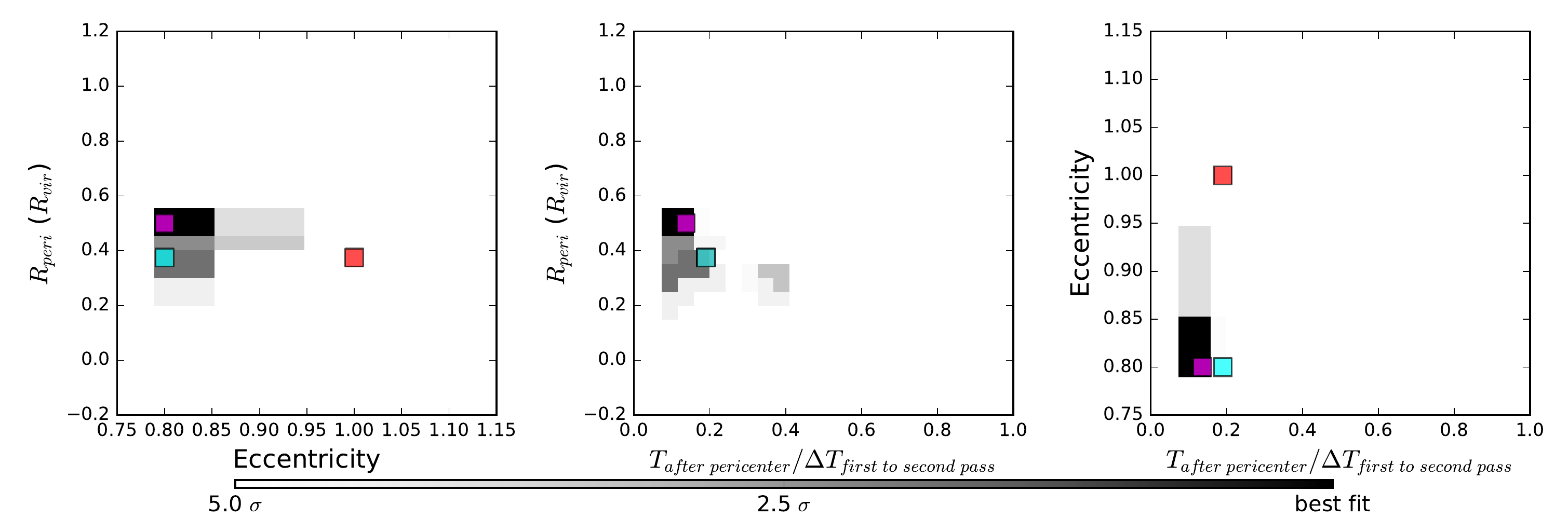}}
\caption[The average score map for modeling \ha\ and HI]
{(a) The average score map for modeling \ha\ kinematics of the Mice galaxy merger. Score map
is a 5-dimensional parameter space. Here we are showing 3 slices of the score map through the best-fit
parameter, across eccentricity, pericentric separation, and time since pericenter. The dark areas
indicate better scores relative to light areas. The scale in the color bars is from average score at the best-fit model (the darkest) down to a score that is lower by five time the standard deviation of scores at the best-fit model (white). The best-fit model is shown with the cyan boxes, so they are located on the darkest points in all panels of (a). The magenta
boxes show the the parameters of the best-fit model of HI. They are located on the darkest point of all panels in (b), which are similar slices of the average score map for modeling HI kinematics of the Mice. Cyan boxes have the same definition in (b). The red boxes in both (a) and (b)
show the best-fit parameters of \cite{Privon:2013fs} which used visual matching to HI kinematics. In the middle panels, the red box
 is hidden behind the cyan box, meaning that \cite{Privon:2013fs} give the same best-fit R$_\textrm{peri}$ and time as our \ha\ model.}
\label{fig:bestfitcut}
\end{figure*}

In Figure \ref{fig:bestfitcut} one can see 3 slices of the 5 dimensional score map at the best-fit 
parameter point, across eccentricity, pericentric separation, and merger stage. In these plots the 
best-fit model can be compared with the best-fit model by \cite{Privon:2013fs}. 

The best-fit parameters for \ha\ and HI kinematics are presented in Table \ref{tab:results}. 
The best-fit eccentricity
is $\leq0.80 \pm^{0.05}_{---}$ and $\leq0.80 \pm^{0.15}_{---}$ in modeling \ha\ and HI kinematics, respectively. 
The reason for the $\leq$ sign is that the range of eccentricities explored in this work was from 0.8 to 1.1.
Obtaining the best eccentricity on the edge of our search range suggests that the reconstructed eccentricity is less than or equal
to this value.
The best-fit pericentric separation is 
$17 \pm^{12}_{6}$ kpc for \ha\ modeling, and is $24 \pm^{8}_{13}$ kpc for
HI modeling. The merger stage defined as the time since pericenter divided by the time from
first passage to second passage, is $0.19\pm^{0.20}_{0.01} $ 
and $0.14\pm^{0.70}_{0.01} $ for \ha\ and HI modelings, respectively. The best-fit time since pericenter
is $200\pm^{40}_{30} $ Myrs for \ha\ and $170\pm^{90}_{50} $ Myrs for HI. The best viewing angles in 
spherical coordinates ($\theta$, $\phi$) relative to the orbital plane is 
$(76^\circ\pm^{3^\circ}_{9^\circ}, 0^\circ\pm^{61^\circ}_{41^\circ})$ 
and $(67^\circ\pm^{4^\circ}_{9^\circ}, 52^\circ\pm^{9^\circ}_{66^\circ})$ for \ha\ and HI models, respectively.
For each Identikit model, we viewed the system from 320 equally separated viewing angles, so the 
resolution of our search in viewing angle was about $11^\circ$. 
The initial orientation $(i,\omega)$ of NGC 4676A is found to be
$(21^\circ\pm^{11^\circ}_{5^\circ}, 29^\circ\pm^{33^\circ}_{38^\circ})$ and 
$(56^\circ\pm^{31^\circ}_{24^\circ}, 50^\circ\pm^{55^\circ}_{96^\circ})$ in modeling 
\ha\ and HI kinematics, respectively. For initial orientation of NGC 4676B we find 
$(38^\circ\pm^{21^\circ}_{10^\circ}, 187^\circ\pm^{20^\circ}_{24^\circ})$ and 
$(69^\circ\pm^{23^\circ}_{27^\circ}, 162^\circ\pm^{25^\circ}_{61^\circ})$ for \ha\ and HI modeling.

\begin{figure*}
\includegraphics[width=0.90\textwidth]{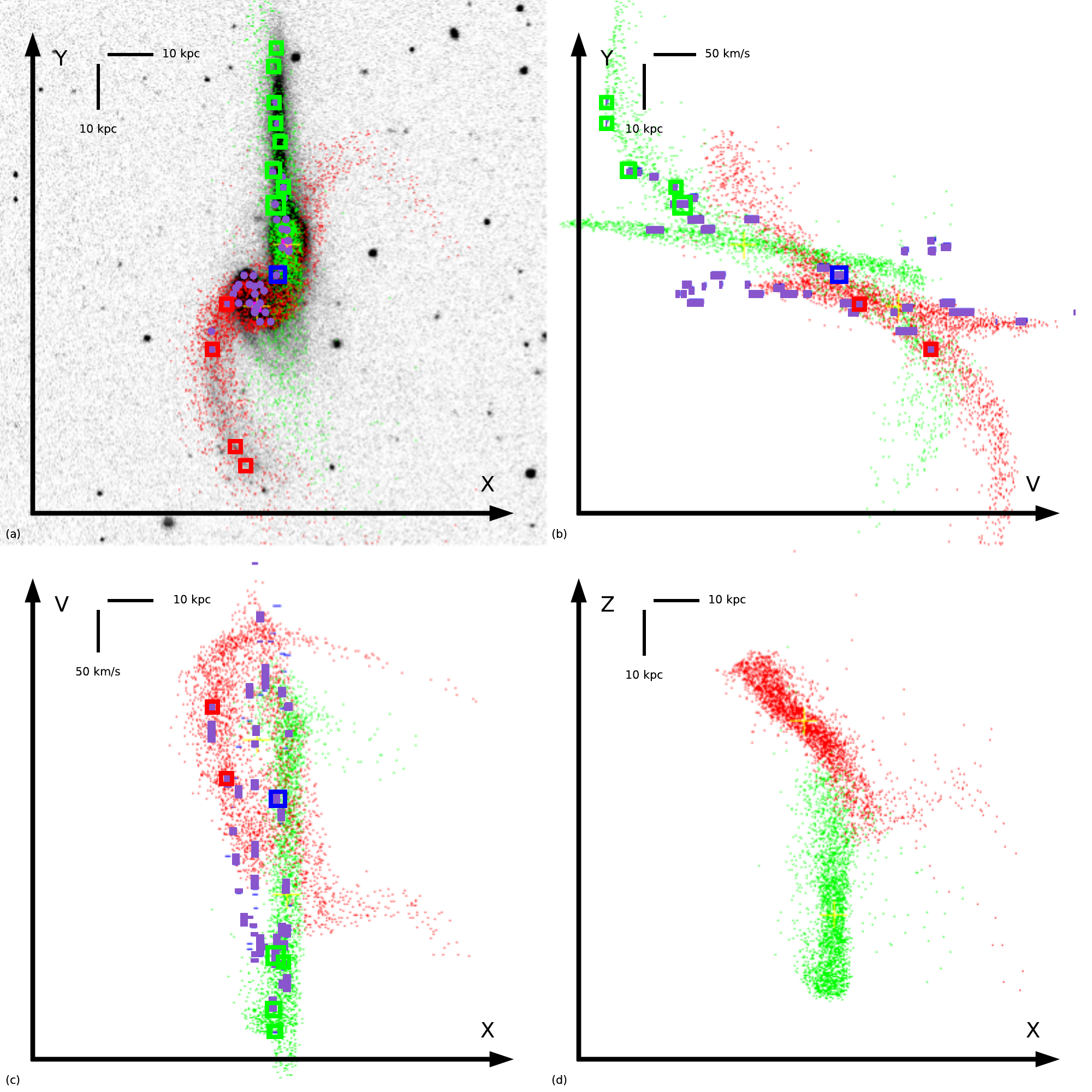}
\caption[The reconstructed model in this work]
{(a) The reconstructed Identikit model (green and red test particles) over-plotted on the SDSS r-band image of the Mice galaxies. The green, red and blue boxes
indicate the phase space regions associated with NGC4676A, NGC4676B, and both
discs, respectively. They are utilized by Identikit to calculate score as a measure of the goodness of 
the match. The purple circles show the position of SparsePak fibers used to measure
the \ha\ kinematics. The Identikit model reproduce the main tidal features of the Mice. However, in the model of the southern galaxy (red particles) we see an extra tidal feature in the north-west direction, which is absent in both HI and optical images. This extra feature is also present in the model presented by \cite{Privon:2013fs}.(b)(c) The line of sight velocity of the reconstructed
model versus the y (x) spatial axis. The position of purple rectangles 
show the \ha\ velocity measured at each fiber, and their extents indicate the dispersion of
the \ha\ emission lines. In panels (b) and (c), the position of red, green, and blue boxes along the V axis show the velocity of phase space regions,
which are determined by the \ha\ velocities (purple rectangles). Some boxes in panel (a) are placed where
there is no \ha\ velocity information (no fiber), therefore they do not have counterparts in panels
(b) and (c), and are only used to constrain the morphology of the model.
(d) The model viewed from the direction of the y axis in panel (a). This
figure can be compared to Figure 5 of \cite{Privon:2013fs}.}
\label{fig:model}
\end{figure*}

All of the reconstructed parameters for modeling HI and \ha\ kinematics are consistent
within $1\sigma$ uncertainty. This can be seen by comparing Figure \ref{fig:bestfitcut}
(a) and (b) in which the dark areas with high scores mostly overlap. Figure
\ref{fig:bestfitcut} also shows the model by \cite{Privon:2013fs} which was 
obtained with visual matching of HI kinematics of the Mice to Identikit models and
confirming the model with a fully self-consistent N-body simulation. \cite{Privon:2013fs} 
only explored models
with parabolic orbits (eccentricity=1). Except eccentricity, our parameters
for modeling \ha\ kinematics are consistent to within $1\sigma$ of the \cite{Privon:2013fs} parameters. 
The parameters of our HI model
are a bit further away, but still within $2\sigma$ of \cite{Privon:2013fs}. Figure \ref{fig:model}
shows one of the best-fit models to the \ha\ kinematics. This figure can be compared to 
Figure 5 of \cite{Privon:2013fs}.

\section{Discussion}
\label{sec:discussion}

The peculiar morphology of the Mice has made it a curious case for dynamical modeling. 
There has been several attempts to model the initial conditions of this system. 
The reported parameters in some of these attempts are shown in Table \ref{tab:previousmodels}.  
In all but one of these dynamical models the mass ratio is presumed to be 1. This assumption is made 
because of the similar stellar mass ratio (1.5:1, \citealt{Wild:2014do}) and the strong tidal tails in both galaxies 
(\citealt{Toomre:1972jia}; \citealt{Barnes:2016gg}).
We also adopt equal mass merger models in this work. 

\begin{table*}
\centering
\begin{tabular}{llllll}
	\hline
	dynamical model								&	$\mu$					&$e$						& $\mathrm{R_{peri}}$	&	physical time since 	&	kinematics					\\ 
															&							&							&	(kpc)							&	pericenter (Myrs)		&										\\ \hline
	\cite{Toomre:1972jia}$^1$			&	1.0					&0.6						&	...								&	120 							&	\ha\ only centers	\\
	\cite{1993ApJ...418...82M}$^2$			&	1.0					&0.6						& 	23								&	180							&	\ha\ long slit		\\
	\cite{Barnes:2004kp}$^3$	&	1.0					&1.0						&	8.9							&	170							&	HI map$^*$					\\
	\cite{Privon:2013fs}$^2$					&	1.0					&1.0						&	14.8							&	175							&	HI map$^*$					\\
	\cite{Holincheck:2016fy}$^1$	&  0.59$\pm$0.52	&0.70$\pm$0.27	&  15.9$\pm$5.0			&  430$\pm$190			& no velocity used				\\
\end{tabular}
\caption[Previous dynamical models of the Mice.]
{Previous dynamical models of the Mice. $^*$ From \cite{Hibbard:1996do}. $^1$ Test particles with central potential. 
$^2$ Collisionless self consistent simulations. $^3$ Hydro simulations including gas and star formation.}
\label{tab:previousmodels}
\end{table*}

Time is the best-constrained encounter parameter for the Mice galaxies so far.
Among the reported values in Table \ref{tab:previousmodels}, 
the lowest value for time since pericenter is 120 Myr for \cite{Toomre:1972jia}. 
The largest is 430 $\pm{190}$ Myr for \cite{Holincheck:2016fy},
and the median is 175 Myr for \cite{Privon:2013fs}. Our reconstructions
from HI and \ha\ modeling are both consistent with the median of 
these models.
In both of our models the Mice is found to be at a relatively early stage. 
The galaxies are currently at about 1/5 (1/8) of time from the first passage to
the second passage for the \ha\ (HI) model. Taking the uncertainty into
account, both values are consistent with \cite{Privon:2013fs} ($\approx$ 1/5).
Our \ha\ and HI models predict that the second passage will occur in $530\pm^{480}_{220}$ Myrs
and $530\pm^{370}_{220}$ Myrs, respectively, also consistent with the model of \cite{Privon:2013fs} for which
the second passage will happen in $\approx$ 430 Myrs.

Both of our models prefer elliptical orbits, which is
consistent with \cite{Toomre:1972jia} and \cite{Holincheck:2016fy}.
Nonetheless, in \cite{Mortazavi:2016hv} it was shown that 
our modeling method underestimates 
the eccentricity by $\sim$0.1 in most tests on hydrodynamical merger simulations
with parabolic orbits.
The large error-bars in the reconstructed eccentricity in 
\cite{Mortazavi:2016hv} indicate that our modeling method is 
not sensitive enough to eccentricity, and other constraints
may be required to better constrain the initial energy of orbits in 
galactic encounters.

Physical pericentric distance, $\mathrm{R_{peri}}$, in our \ha\ 
and HI models are $17 \pm^{12}_{6}$ kpc and $24 \pm^{8}_{13}$ kpc which are consistent
with all previous reported values except that of \cite{Barnes:2004kp} (8.9 kpc). 
$\mathrm{R_{peri}}$ is hard to constrain, as its effect on morphology of tides
is correlated with mass profile of isolated galaxies, 
i.e. disc scale-length and halo scale-radius, as well as the luminous 
mass fraction\footnote{Luminous mass fraction $f_L=\frac{M_d+M_b}{M_d+M_b+M_h}$. In
Identikit models used in this work $f_L=0.2$.} \citep{Barnes:2016gg}.
In \cite{Mortazavi:2016hv} we showed that our method overestimates
$\mathrm{R_{peri}}$ in most of the tests on hydrodynamical simulations, suggesting
that our reconstruction of $\mathrm{R_{peri}}$ of the Mice is an upper limit.


Most recently, \cite{Holincheck:2016fy} used judgment of citizen scientists
(Galaxy Zoo, \citealt{Lintott:2010gb}) to match the morphology of a relatively large sample
of merger systems. In addition to parameters explored in this work, they varied mass ratio
parameter, and they managed to present best-fit
models for a relatively large sample of 62 tidally interacting galaxy pairs. For each galaxy pair 
$\sim 4.6\times 10^6$ parameter space points were rejected initially for not displaying
tidal features, and $\sim$ 50,000 parameter space points were
viewed and judged visually by citizen scientists. 
However,  as seen in Table \ref{tab:previousmodels} , 
the uncertainty of their measurements are relatively large compared 
to the errors in this work. 
Lack of kinematic information in their analysis and fewer parameter space points contribute to larger error bars. In this work, we calculate the score for $3.2\times10^{10}$ points
in the parameter space elaborated in Table \ref{tab:Identikit}.

The strong tidal tails and the edge-on view of the Mice makes it a straight-forward case for 
testing a galaxy merger model. However, one may ask how common the encounter parameters
of the Mice are among galaxy mergers. While the answer to this question requires modeling a larger
sample of mergers, we can compare our reconstructed encounter parameters of the Mice, with the distribution of orbital parameters of
dark matter halo mergers in cosmological simulations. The eccentricity of the Mice found in this work ($e\leq0.80$), 
is lower than the most frequent
eccentricity ($e=1.0$) among dark matter halo mergers (e.g. see Figure 11 of \citealt{Benson:2005hi}
and Figure 6 of \citealt{Khochfar:2006de}). Parabolic ($e=1$) orbits have zero total mechanical 
energy, and are expected to be seen if the protogalaxies are initially at rest and far from each other. 
The uncommonly low eccentricity we found may be due to a bias in our modeling method.
On the other hand, it may indicate that orbits of interacting galaxies decay earlier than
the dark matter halos they live in.
The pericentric separation found for the Mice ($0.38\pm^{0.16}_{0.09}\ \mathrm{R_{vir}}$ and 
$0.50\pm^{0.03}_{0.44}\ \mathrm{R_{vir}}$ for \ha\ and HI, respectively)
 are both uncertain. They both  
 fall into the most common values in the distribution of $\mathrm{R_{peri}}$s in dark 
matter halo mergers of the Millennium cosmological simulation (see Figure 4 
of \citealt{Khochfar:2006de}). A sanity check for this answer is to calculate the 
period for this elliptical orbit to see if it is longer than the Hubble time. Otherwise,
one may ask why the galaxies did not merge during the previous pericenter. Assuming that the 
mass of the dark matter halo of the larger partner is $10^{11}-10^{12}M_\odot$ the period of 
the orbit will be 10-3 Gyrs. This is less than the age of the Universe and suggests that the
orbit has become bound due to recent processes in the last 3-10 Gyrs, e.g.
recent mass growth due to minor mergers.

Merger induced star formation 
rate (SFR) is affected by the encounter parameters in hydrodynamical simulations. 
\cite{Cox:2008jj} show that 
mergers with larger pericentric distance induce starbursts at later times. Moreover, 
the relative inclination of the discs with respect to 
orbital plane correlates with the total amount 
of merger induced star formations (also see \citealt{Snyder:2011fs}). 
On the other hand, the star formation in these simulations 
depends on presumptions about sub-grid physics, which are the physical processes
that happen in scales smaller than the resolution of the simulation. Even the most
recent simulations of galaxy formation do not yet resolve particles at
scales smaller than stars. To implement star formation, most current
simulations assume a density threshold above which stars are formed.
The exact value of this threshold is different in different simulations and 
varies with the resolution of the hydrodynamical simulation (e.g. see \citealt{Brooks:2015up}),
and they are usually calibrated to reproduce the Kennicutt-Schmidt relationship 
between gas mass surface density and star formation rate \citep{Kennicutt:1998ki}. 
\cite{Barnes:2004kp} proposed a shock-induced star formation recipe in order to reproduce 
extended merger induced star formation
observed in tidal tails of interacting galaxies (e.g. see \citealt{deGrijs:2003jr}). 
In addition to sub-grid star formation recipe, the model for feedback 
from SNe, stellar winds, and AGN adds to the complications of
star formation model in hydrodynamical simulations 
(e.g. see \citealt{Cox:2006kba,Oppenheimer:2010ic}). Finding independent
constraints on star formation history in galaxy mergers improves our
understanding of the sub-grid physics in hydrodynamical simulations. 

Our estimate of the merger stage and other encounter parameters of 
the Mice is independent of star formation rate. It only depends on the 
dynamics of tidal features. As a result, by looking at a ``Mice-like" simulation 
of a galaxy merger with the same encounter parameters and at the same merger stage that we found in this work, we can put an independent constraint on 
the sub-grid physics of merger induced star formation. 
To do so, we can compare the total enhancement of star formation 
in the hydrodynamical simulation to the measured star formation in the Mice.
Unfortunately, our data is not flux calibrated and we can not use it to estimate current star formation rate in the Mice galaxies. On the other hand, other measurements of the SFR in the Mice is uncertain and varies based on 
the observed star formation indicator. \cite{Wild:2014do} provide different star formation
measurements for the Mice combining CALIFA IFU spectroscopy with other archival data. 
In particular, the SFR derived from ionized gas 
recombination lines (\ha\ and H$\beta$) is significantly higher than the one 
derived from modeling the stellar continuum. While the difference may be an 
indicator of very recent star formation, in \S \ref{sec:shockedregions} 
we showed that about 23\% of the \ha\ line emission arises from sources other than 
photoionization. In addition, large amounts of dust attenuation, particularly
at the centers of the galaxies, introduce large uncertainties in star formation rate measurements. 
The edge-on view of the discs may have played a role in worsening this problem. However, this works demonstrates that improvements in the measurement of star formation and dynamical merger stage of 
the Mice (or any other merger system) provides a powerful tool for testing 
sub-grid physics in hydrodynamical simulations.

\subsection{Summary}

In this work we modeled the initial conditions of the Mice galaxy merger 
system, utilizing an automated method based on
Identikit software package. We observed the Mice with the SparsePak IFU
 on the WIYN telescope at KPNO, considered one- and 
two-component emission lines, and separated the emission from
 photo-ionized regions and shocked regions
using a plot of \nii/\ha\ vs. velocity dispersion. 

We used the kinematics of both photo-ionized 
gas (\ha) and cold gas (HI) to 
compare the effect of using different kinematic tracers. 
We found consistent results for the two 
kinematic tracers. They are also consistent with 
previous models for the Mice, particularly that of \cite{Privon:2013fs}. 
This work suggests that while kinematic information on tidal features is still needed, 
our automated method of dynamical modeling is 
applicable to some major 
galaxy merger systems using both HI and 
\ha\ velocity information. 
We can use data from
IFU surveys (e.g. MaNGA, SAMI, etc. ) and high resolution HI surveys (e.g. MeerKAT) 
to model the dynamics of major galaxy mergers.
Though, one should be cautious that in some mergers systems
gas does not follow stars on large scales due to dissipative processes like
ram pressure striping. 
Dynamical reconstruction of merger 
stage can put independent 
constraints on the sub-grid physics of star formation 
in hydrodynamical simulations of galaxies. At this point, this is not 
possible for the Mice system, mostly because of the large uncertainty in
star formation rate measurements.

\section*{Acknowledgments}

We would like to thank the referee, because her/his suggestions improved this paper significantly. This project was supported in part by the STScI DDRF. This work 
used the Extreme Science and Engineering Discovery Environment 
(XSEDE), which is supported by National Science Foundation grant number ACI-1053575 
(see \citealt{Towns2014XSEDE:Discovery}). G.C.P. was supported by a FONDECYT 
Postdoctoral Fellowship (No. 3150361).




\bibliographystyle{mnras}
\bibliography{Mendeley} 








\bsp	
\label{lastpage}
\end{document}